\documentclass[pra,aps,floatfix,floats,superscriptaddress,notitlepage, twocolumn]{revtex4-2} 
\usepackage{amsfonts,bbm,bm,braket, comment}
\usepackage{scalerel}
\usepackage{enumerate}
\usepackage{mathtools}
\usepackage{ifthen}
\usepackage{tensor}
\usepackage{tikz}
\usepackage{tikz-network}
\usetikzlibrary{patterns,decorations.pathreplacing}
\usepackage[colorlinks,bookmarks=false,citecolor=NavyBlue,linkcolor=OliveGreen,url color=blue]{hyperref}

\usepackage{color}
\definecolor{myred}{RGB}{232,102,102}
\definecolor{myblue}{RGB}{187,187,255}
\definecolor{myorange}{RGB}{202,52,51}
\definecolor{mygrey}{RGB}{105,105,105}
\definecolor{OliveGreen}{RGB}{85,107,47}
\definecolor{NavyBlue}{RGB}{0,0,128}
\definecolor{mygreen}{RGB}{34,139,34}
\definecolor{myY}{RGB}{220,255,203}
\definecolor{myYO}{RGB}{255, 220, 151}
\definecolor{myblue1}{RGB}{176,223,229}
\definecolor{myblue2}{RGB}{0,0,128}
\definecolor{myblue3}{RGB}{0,108,255}
\definecolor{myblue4}{RGB}{101,147,245}
\definecolor{myblue5}{RGB}{115,194,251}
\definecolor{myblue6}{RGB}{87,160,211}
\definecolor{myblue7}{RGB}{137,207,240}
\definecolor{myblue8}{RGB}{29,41,81}
\definecolor{myblue9}{RGB}{14,77,146}
\definecolor{myblue10}{RGB}{15,82,186}
\definecolor{myred1}{RGB}{255,36,0}
\definecolor{myred2}{RGB}{205,92,92}
\definecolor{myred3}{RGB}{178,34,34}
\definecolor{myred4}{RGB}{164,90,82}
\definecolor{myred5}{RGB}{255,8,0}
\definecolor{myred6}{RGB}{202,52,51}
\definecolor{myred7}{RGB}{66,13,9}
\definecolor{myred8}{RGB}{141,2,31}
\definecolor{myred9}{RGB}{250,128,114}
\definecolor{myred10}{RGB}{237,41,57}
\definecolor{myyellow1}{RGB}{254,220,86}
\definecolor{myyellow2}{RGB}{255,229,180}
\definecolor{myyellow3}{RGB}{238,220,130}
\definecolor{myyellow4}{RGB}{253,165,15}
\definecolor{myyellow5}{RGB}{255,195,11}
\definecolor{myyellow6}{RGB}{218,165,32}
\definecolor{myyellow7}{RGB}{255,211,0}
\definecolor{myyellow8}{RGB}{248,222,126}
\definecolor{myyellow9}{RGB}{245,245,220}
\definecolor{myyellow10}{RGB}{248,228,115}

\definecolor{mygray1}{RGB}{246,246,246}
\definecolor{mygray2}{RGB}{32,32,32}
\definecolor{mygray3}{RGB}{64,64,64}
\definecolor{mygray4}{RGB}{96,96,96}
\definecolor{mygray5}{RGB}{128,128,128}
\definecolor{mygray6}{RGB}{160,160,160}
\definecolor{mygray7}{RGB}{224,224,224}
\definecolor{mygray8}{RGB}{180,180,180}

\usepackage{tensor}
\newcommand{\be}{\begin{equation}}
\newcommand{\ee}{\end{equation}}
\newcommand{\bea}{\begin{eqnarray}}
\newcommand{\eea}{\end{eqnarray}}
\newcommand{\ba}{\begin{aligned}}
\newcommand{\ea}{\end{aligned}}
\newcommand{\bw}{\begin{widetext}}
\newcommand{\ew}{\end{widetext}}
\newcommand{\1}{\mathbbm{1}}

\newcommand{\Wgategreen}[2]{
\draw[very thick] (#1-0.5, #2 +0.5) -- (#1+0.5,#2-0.5);
\draw[very thick] (#1-0.5,#2-0.5) -- (#1+0.5,#2+0.5);
\draw[ thick, fill=mygreen, rounded corners=2pt] (#1-0.25,#2+0.25) rectangle (#1+0.25,#2-0.25);
\draw[thick] (#1,#2+0.15) -- (#1+0.15,#2+0.15) -- (#1+0.15,#2);
}

\newcommand{\Wgateolivegreen}[2]{
\draw[very thick] (#1-0.5, #2 +0.5) -- (#1+0.5,#2-0.5);
\draw[very thick] (#1-0.5,#2-0.5) -- (#1+0.5,#2+0.5);
\draw[ thick, fill=OliveGreen, rounded corners=2pt] (#1-0.25,#2+0.25) rectangle (#1+0.25,#2-0.25);
\draw[thick] (#1,#2+0.15) -- (#1+0.15,#2+0.15) -- (#1+0.15,#2);
}

\hypersetup{
pdftitle={Exact Spectral Statistics in Strongly localised Circuits},
pdfsubject={Quantum chaos},
pdfauthor={Bruno Bertini, Pavel Kos, and Tomaz Prosen},
pdfkeywords={quantum chaos, quantum many-body systems}
}

\begin{document}

\title{Exact Spectral Statistics in Strongly Localised Circuits}

\author{Bruno Bertini}
\affiliation{Rudolf Peierls Centre for Theoretical Physics, Clarendon Laboratory, Oxford University, Parks Road, Oxford OX1 3PU, United Kingdom}
\affiliation{School of Physics and Astronomy, University of Nottingham, Nottingham, NG7 2RD, United Kingdom}

\author{Pavel Kos}
\affiliation{Department of Physics, Faculty of Mathematics and Physics, University of Ljubljana, Jadranska 19, SI-1000 Ljubljana, Slovenia}
\affiliation{Max-Planck-Institut fur Quantenoptik, D-85748 Garching, Germany}

\author{Toma\v z Prosen}
\affiliation{Department of Physics, Faculty of Mathematics and Physics, University of Ljubljana, Jadranska 19, SI-1000 Ljubljana, Slovenia}

\begin{abstract}

Since the seminal work of Anderson, localisation has been recognised as a standard mechanism allowing quantum many-body systems to escape ergodicity. This idea acquired even more prominence in the last decade as it has been argued that localisation --- dubbed many-body localisation (MBL) in this context --- can sometimes survive local interactions in the presence of sufficiently strong disorder.  A conventional signature of localisation is in the statistical properties of the spectrum --- spectral statistics --- which differ qualitatively from those in the ergodic phase. Although features of the spectral statistics are routinely used as numerical diagnostics for localisation, they have never been derived from first principles in the presence of non-trivial interactions. Here we fill this gap and provide the example of a simple class of quantum many-body systems --- which we dub strongly localised quantum circuits --- that are interacting, localised, and where the spectral statistics can be characterised exactly. Furthermore, we show that these systems exhibit a cascade of three different regimes for spectral correlations depending on the energy scale: at small, intermediate, and large scales they behave as disconnected patches of three decreasing sizes. We argue that these features appear in generic MBL systems, with the difference that only at the smallest scale they do become Poissonian.

\end{abstract}

\maketitle

 \section{Introduction}

Analysing the statistical properties of the spectrum of the time-evolution operator is one of the most universal and versatile routes to quantify ergodicity and quantum chaos in interacting quantum systems~\cite{haake2018}. This approach can be applied to both autonomous and periodically driven systems and is based upon two fundamental conjectures: the quantum chaos (aka Bohigas) conjecture~\cite{bohigas1984characterization,casati1980on,berry1981quantizing} and the Berry-Tabor conjecture~\cite{berry1977level}. The former asserts that the spectral statistics of few-particle quantum systems with chaotic classical limits is equivalent to that of an ensemble of random matrices with the same anti unitary symmetries~\cite{mehta2014random}. The latter states that the spectrum of few-particle integrable systems is equivalent to a collection of uncorrelated Poissonian events on the energy line (at least over small enough intervals).

Both these classic conjectures have been extended to the many-body realm: The emergence of random matrix theory (RMT) spectral statistics is considered the hallmark of quantum ergodicity. In contrast, the emergence of Poissonian level statistics signals localisation, which may emerge either due to Bethe-ansatz integrability or quenched disorder. Despite an abundance of numerical evidence, the first rigorous results establishing RMT spectral statistics for some classes of periodically driven quantum lattice systems, the so-called dual-unitary circuits~\cite{bertini2019exact}, appeared only recently~\cite{PRL2018,CMP2021}. On the other hand, Poissonian level statistics has been rigorously connected to Anderson localisation in non-interacting disordered systems~\cite{wang2001localization,aizenman}, and its emergence in disordered systems with local interactions has been considered a hallmark for the putative many-body localisation (MBL), which has been inferred to emerge in both autonomous and Floquet systems~\cite{abanin2019many,nandkishore2015many}. Despite this being a highly investigated subject there is currently no known example of interacting many-body quantum systems, for which Poissonian, or more generally, non-RMT spectral statistics can be established exactly. The purpose of this paper is to fill this gap: we formulate a class of locally-interacting and localised Floquet systems that closely resemble the standard systems exhibiting Floquet MBL~\cite{chan2018spectral, sunderhauf2018localization, garratt2021many, garratt2021local2, FloquetMBL21} but for which the local integrals of motion (LIOM) can be constructed explicitly, and the spectral statistics can be characterised exactly. In particular, we show that in strongly localised circuits the computation of all moments of the spectral form factor (SFF), i.e.\ the Fourier transform of the spectral density's two-point function, can be reduced to a study of a finite-dimensional spatial (dual) transfer matrix. We then obtain closed-form results for the disorder-averaged SFF, which agree with the Poissonian statistics, as well as asymptotic results for the higher moments of the SFF, which deviate from Poisson. We argue that such deviations of higher-point statistics from the Poissonian result are ubiquitous for systems in the MBL regime if one looks at correlations among energy levels that are distant enough, i.e. at large ``energy scales". In contrast with our solvable example, however, generic MBL systems appear to show Poissonian higher moments if one looks at correlations between energy levels that are sufficiently close.

The rest of the paper is organised as follows. In Sec.~\ref{sec:setting} we describe the setting considered and introduce strongly localised circuits. In Sec.~\ref{sec:spectralstat} we analyse the spectral statistics of these circuits. In Sec.~\ref{sec:MBL} we compare our exact results with those obtained in generic finite-size MBL systems. Finally, Sec.~\ref{sec:conclusions} reports our conclusions. Some of the more technical aspects of our analysis are reported in the appendices.

\section{Setting}
\label{sec:setting}

We formulate our minimal models in the framework of brickwork quantum circuits, which are known to provide useful idealisations of quantum many-body systems with local interactions~\cite{nahum2017quantum, chan2018solution, vonkeyserlingk2018operator, khemani2018operator} and, in particular, are now the standard framework to study spectral statistics in the many-body realm~\cite{chan2018spectral, sunderhauf2018localization, garratt2021many, PRL2018, CMP2021, friedman2019spectral, flack2020statistics, mouldgalya2021spectral,garratt2021local, garratt2021local2, chan2021manybody, kos2021thermalisation}. We consider a set of $2L$ qubits that are positioned along a periodic lattice of half-integers $\mathbb Z_{2L}/2$
and are evolved in time by discrete applications of a unitary operator over $\mathcal H=(\mathbb C^2)^{\otimes 2L} : $  
\be 
\mathbb U_{L} = \prod_{x \in \mathbb Z_{L}} \eta_{x}(W_{x}) \!\!\!\!\prod_{x \in \mathbb Z_{L}+\frac{1}{2}}\!\!\!\!\eta_{x}(U_{x}) \,. 
\label{eq:Floquet}
\ee
Here we introduced the ``local gates'' $U_{x}, W_{x}$, i.e.\ a set of $2L$ four dimensional unitary matrices specifying the interactions among two neighbouring qubits, and the ``positioning operator" $\eta_{x}(\cdot)$. This is a linear map that places a generic local operator $O$ on the $2L$-qubit chain in such a way that its right edge is at position $x$.

In the following we consider a family of brickwork quantum circuits with local gates of the form
\be
U_{x} = (u_{x-\frac{1}{2}} \otimes \1 )\, U, \qquad  W_{x} =( \1 \otimes w_{x})\, W,
\label{eq:Floquetgates}
\ee
where $u_{x}, w_{x}$ are position-dependent unitary matrices (without loss of generality we take them in ${\rm SU}(2)$) representing the static disorder, while   
\be
U= e^{i J_u Z\otimes Z}, \quad W= e^{i J_w Z\otimes Z},\quad J_{u},J_w\in[0,\pi/4],
\label{eq:SL}
\ee 
model non-trivial interactions. Here  $\{X,Y,Z\}$ indicate the triple of Pauli matrices. 

The circuits (\ref{eq:Floquet},\ref{eq:Floquetgates},\ref{eq:SL}) are similar to several Floquet MBL quantum circuits studied in the recent literature~\cite{chan2018spectral, sunderhauf2018localization, garratt2021many, FloquetMBL21, garratt2021local2}, their only special feature is that the disorder is applied every two sites rather than at each site. This seemingly minor difference, however, turns out to lead to \emph{exact solvability}. To see how this happens we note that these circuits preserve the projectors on eigenstates $\ket{\pm}$ of $Z$, $Z\ket{\pm}=\pm\ket{\pm}$,
that are placed at half-odd integer sites
\be
\mathbb U_L^\dag\cdot \prod_{j=1}^n \eta_{x_j+{\scriptscriptstyle \frac{1}{2}}}(\ket{\pm}\!\!\bra{\pm})\cdot \mathbb U_L = \prod_{j=1}^n \eta_{x_j +{\scriptscriptstyle \frac{1}{2}}}(\ket{\pm}\!\!\bra{\pm})\,,
\ee
where ${x_j}\in\mathbb Z_L$. This property implies that any one-qubit operator $O$ surrounded by $\ket{\pm}\!\bra{\pm}$ cannot spread in time. In particular 
\begin{align}
&\mathbb U_L^\dag \eta_{x+{\scriptscriptstyle \frac{1}{2}}}\!(\ket{\mu}\!\!\bra{\mu}\!\otimes\! O\! \otimes\! \ket{\nu}\!\!\bra{\nu})\mathbb U_L \notag\\
&\qquad\qquad\qquad= \eta_{x+{\scriptscriptstyle \frac{1}{2}}}\!(\ket{\mu}\!\!\bra{\mu}\!\otimes\!v^{\dag}_{\mu\nu,x}Ov^{\phantom{\dag}}_{\mu\nu,x}\!\!\otimes\! \ket{\nu}\!\!\bra{\nu}),
\end{align}
where $\mu,\nu=\pm$ and we defined 
\be
v_{\mu\nu,x}=w_x e^{i \mu J_w Z} u_x  e^{i \nu J_u Z}\,.
\label{eq:vdef}
\ee
Therefore one can easily find conserved quantities. Specifically, introducing the local operators  
\be
I^{(\mu\nu\iota)}_x\!= \eta_{x+{\scriptscriptstyle \frac{1}{2}}}(\ket{\mu}\!\!\bra{\mu} \!\otimes\! \ket{\iota\phi_{\mu\nu,x}}\!\!\bra{ \iota\phi_{\mu\nu,x}}\!\otimes\! \ket{\nu}\!\!\bra{\nu}),
\label{eq:LIOM}
\ee
where $\ket{\pm\phi_{\mu\nu,x}}$ are the two eigenstates of $v_{\mu\nu,x}$, we see that they form an extensive set of local integrals of motion, i.e.,
\be
\!\!\!I^{(\mu\nu\iota)}_x=\mathbb U_L^\dag I^{(\mu\nu\iota)}_x\mathbb U_L, \qquad x\in\mathbb Z_L,\quad \mu,\nu,\iota=\pm\,.
\label{eq:LIOMprop}
\ee
It is easy to see that, in addition, the operators in \eqref{eq:LIOM} commute~\footnote{The commutativity of $\{I^{(\mu \nu\iota)}_x\}$ can be shown in two steps. First one notes that $I_x^{(\mu\nu\iota)}$ and $I_{y\neq x}^{(\mu'\nu'\iota')}$ can only overlap on a site with commuting
projectors
$\ket{\pm}\bra{\pm}$, hence they commute. Next one observes ${I_x^{(\mu\nu\iota)} I_x^{(\mu'\nu'\iota')}= \delta_{\mu,\mu'}\delta_{\nu,\nu'}\delta_{\iota,\iota'}}$, which concludes the proof.}, and their eigenvalues can be used to specify a basis of the Hilbert space. Indeed, taking products of these operators we can construct projectors on all the states of a basis of $\mathcal H$. Explicitly we have  
\be
\!\!\!\!\prod_{x=1}^{L-1}  I_x^{(\mu_x\nu_x\mu_{x+1})} I_L^{(\mu_L\nu_L\mu_{1})} = \ket{\Psi_{\{\mu_x\}, \{\nu_x\}}}\!\!\bra{\Psi_{\{\mu_x\}, \{\nu_x\}}}\!,
\label{eq:Ibasis}
\ee
where we introduced  
\be
\ket{\Psi_{\{\mu_x\}, \{\nu_x\}}}= \bigotimes_{x=1}^L \ket{\mu_x}\otimes\ket{\nu_x \phi_{\mu_x\mu_{x+1},x}}\otimes\ket{\mu_{x+1}},
\ee
and set $\mu_{L+1}=\mu_{1}$. From this expression it is immediate to see that 
\be
\{\ket{\Psi_{\{\mu_x\}, \{\nu_x\}}}\!\!:\,\,  \mu_x =\pm, \,\, \nu_x =\pm,\,\, x=1,\ldots L\},
\label{eq:eigenbasis}
\ee
is indeed a basis of the Hilbert space.

The above properties imply that \eqref{eq:LIOM} is a complete set of LIOM for the circuits (\ref{eq:Floquet},\ref{eq:Floquetgates},\ref{eq:SL}) and, therefore, the latter are many-body localised~\cite{nandkishore2015many, abanin2019many} for any choice of $J_{u}$, $J_w$, $\{u_x\}$ and $\{w_x\}$. In fact, they form a special class of many-body localised systems, which we term  ``strongly localised",  where the LIOM have strictly finite support. In these circuits one can explicitly determine the many-body spectrum of the Floquet operator  \eqref{eq:Floquet} by applying it to the elements of \eqref{eq:eigenbasis}. Indeed, \eqref{eq:Ibasis} and \eqref{eq:LIOMprop} guarantee that \eqref{eq:eigenbasis} is an eigenbasis of $\mathbb U_L$. In particular, direct application reveals 
\be
\mathbb U_L \ket{\Psi_{\{\mu_x\}, \{\nu_x\}}} = \prod_{x=1}^L e^{ i \nu_x \phi_{\mu_x \mu_{x+1},x}}\ket{\Psi_{\{\mu_x\}, \{\nu_x\}}}\,, 
\label{eq:eigenvalues}
\ee
where $e^{\pm i \phi_{\mu\nu,x}}$ are the two eigenvalues of $v_{\mu\nu,x}$ (cf.~\eqref{eq:vdef}). We stress that this discussion applies for any distribution of $u_{x}$ and $w_{x}$. 
 
\section{Spectral Statistics}
\label{sec:spectralstat}
 
To characterise the spectral statistics of strongly localised circuits we consider the SFF and its higher moments, in our setting expressed as~\cite{chan2018spectral,CMP2021}
\be
K_n(t,L)= \mathbb{E}\left[ |{\rm tr}[\mathbb U_{L}^t]|^{2n} \right],\qquad n=1,2,\ldots,
\label{eq:SFFmoments}
\ee
where $\mathbb E[\cdot]$ is an average required to make these quantities well defined (they are generically not self-averaging~\cite{prange1997the}). In the following we take $\mathbb E[\cdot]$ to be the average over random, identically and independently distributed $u_{x}$ and $w_{x}$. Note that $t$ in Eq.~\eqref{eq:SFFmoments} sets the inverse energy difference (scale) at which the spectral correlations are probed.

Using the explicit, locally correlated form of the eigenvalues in Eq.~\eqref{eq:eigenvalues}, the moments of SFF~\eqref{eq:SFFmoments} can be written as a partition sum of a $2n\times L$ statistical model. 
Since $u_{x}, w_{x}$ are identically and independently distributed we can directly rewrite the moments
\be
K_n(t,L) = {\rm tr}[\mathbb B_n^L],
\label{eq:SFFB}
\ee
in terms of  ${4^n\times4^n}$ transfer matrix $\mathbb B_n$ with elements
\be
[\mathbb B_n]_{a,b} =\mathbb{E}\!\left[ \prod_{j=1}^n\! M^{\phantom{\ast}}_{[a]_j [b]_j} M^*_{[a]_{n+j} [b]_{n+j}}\!\right]\!\!.
\label{eq:Tabcd}
\ee
Here $a,b=0,\ldots,4^n-1$, $[x]_j$ is the $j$-th digit of $x$ in base 2, and we introduced 
\be
M_{ab} = \sum_{\mu,\nu =\pm}\!\! \mu^a \nu^b\! \cos(\phi_{\mu\nu}t)\,.
\label{eq:Mabsimp}
\ee
For completeness, in the appendices  we derive the Eq.~\eqref{eq:SFFB} using the ``space-time duality approach" of Refs.~\cite{PRL2018,CMP2021}.

\begin{figure}[t]
    \centering
    \includegraphics[width=0.45\textwidth]{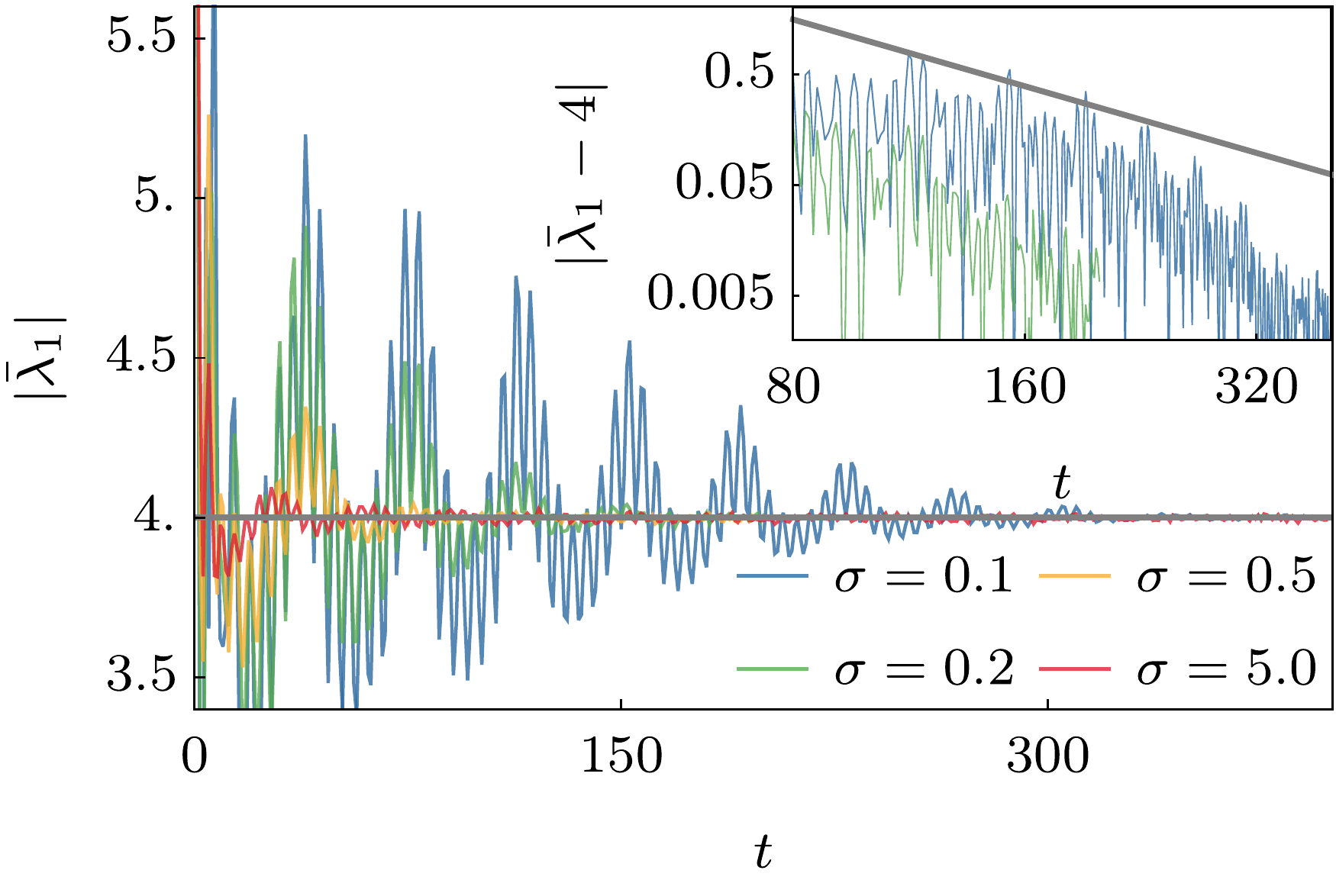}
    \caption{
    Magnitude of the largest eigenvalue of $\mathbb T_1$ versus $t$ for $J_u=0.3$, $J_w=0.4$, and different disorder distributions. We took $u=v=e^{i h}$ where $h$ is taken from GUE(2) with the distribution $P(h)\propto\exp\bigl(-{{\rm tr} (h-h_0)^2}/({2\sigma^2})\bigr)$ and $h_0=0.2 X +1.6 Y + 2.2 Z$. The inset shows $|\bar{\lambda}_1-4|$ and the grey line is $1/t^2$.
    }
    \label{fig:gendistributions}
\end{figure}

Using the expression \eqref{eq:Mabsimp} one can evaluate and diagonalise the matrix $\mathbb B_n$ very efficiently, and in simple cases even analytically. For instance, for ${n=1}$ and Haar distributed  $u, w$ (uniformly on SU(2)) one has ${\rm Sp}[\mathbb B_1]=\{\bar \lambda_1(t), \lambda_1(t),0\}$ where the eigenvalue $0$ is twofold degenerate and we have explicit expression of $\bar \lambda_1(t)$ and $\lambda_1(t)$ (see the appendices). In particular, for large times we have
\be
\bar \lambda_1(t) \simeq 4+4  \frac{\sin (2 J_u t) \sin (2 J_w t)}{\tan (2 J_u) \tan (2 J_w) t^2}\,,
\ee
while $\lambda_1(t) \simeq  O(1/t^2)$. This yields the SFF:
\be
\!\!\!K_1(t,L)\! =\! {\rm tr}[\mathbb B_1^L] \simeq 2^{2L}\! \left[1 + \frac{\sin (2 J_u t) \sin (2 J_w t)L}{\tan (2 J_u) \tan (2 J_w) t^2}\right]\!.
\label{eq:exactK1haar}
\ee
The fact that the infinite time limit of $K_1(t,L)$ gives $2^{2L}$ is very general and does not depend on the specific distribution of $u,w$. This is because the matrix elements of $\mathbb B_n$ are written, because of the averages $\mathbb E[\cdot]$, in terms of integrals of oscillating phases (cf.~\eqref{eq:Mabsimp}). Under very mild regularity conditions on the integrand their infinite-time limit vanishes unless the phases exactly cancel. Therefore we can assume 
\be
\!\!\!\!\!\!\lim_{t\rightarrow \infty}\!\mathbb E[\!\!\prod_{\mu,\nu=\pm}\!\!\!\!\cos(\phi_{\mu\nu}t)^{p_{\mu\nu}}]\!=  \!\! \frac{1}{2^{2n}}\!\!\!\prod_{\mu,\nu=\pm}\!\!\! B(p_{\mu\nu}), \,\, p_{\mu\nu}\!\in\!\mathbb N_0,
\label{eq:infinitetimeav}
\ee
where $B(k)=0$ for $k$ odd and $B(k)=k!/((k/2)!)^2$ for $k$ even. To derive \eqref{eq:infinitetimeav} we expanded cosines in terms of exponentials and picked the constant term. In the case of $\mathbb B_1$ the above formula gives
\be
\lim_{t\rightarrow \infty} \mathbb B_1 \!=\! \begin{bmatrix}
 2 & 0 & 0 &  2 \\
0 & 0 &0 & 0  \\
0 & 0 &0 & 0 \\
2 & 0 & 0 &  2
\end{bmatrix}\quad\Rightarrow\quad \lim_{t\rightarrow \infty}  K_1(t,L)\! = 2^{2L}\!.
\ee
Another general property following from the oscillating-phase-integral form of the matrix elements of $\mathbb B_n$ is that the infinite-time limit is approached in a power-law fashion. In the case of $K_1(t,L)$, our exact result \eqref{eq:exactK1haar} implies that this power is bounded from below by $2$ for all non-singular distributions~\footnote{This is essentially because multiplying an oscillating phase by a non-singular function cannot change the saddle point}. Our numerical experiments confirm this; see Fig.~\ref{fig:gendistributions} for a representative example.

\begin{table}
\begin{tabular}{|| c || c | c | c | c | c | c | c ||}
\hline
$n$  & 1 & 2 & 3 & 4 &5 & 6 & 7 \\ 
\hline
$\bar \lambda_{n}$ & 4 & 24 & 181.193 & 1616.01 & 16318.6 & 180094.0 & 2114717.4 \\
\hline
 $d_n$ & 1 & 1 & 1 & 1 &1 & 1 & 1\\
\hline
\end{tabular}
\caption{Maximal eigenvalue of $\mathbb B_n$ and its degeneracy in the infinite-time limit.}
\label{tab:lambdas}
\end{table}

\begin{figure}[ht]
    \centering
    \includegraphics[width=0.45\textwidth]{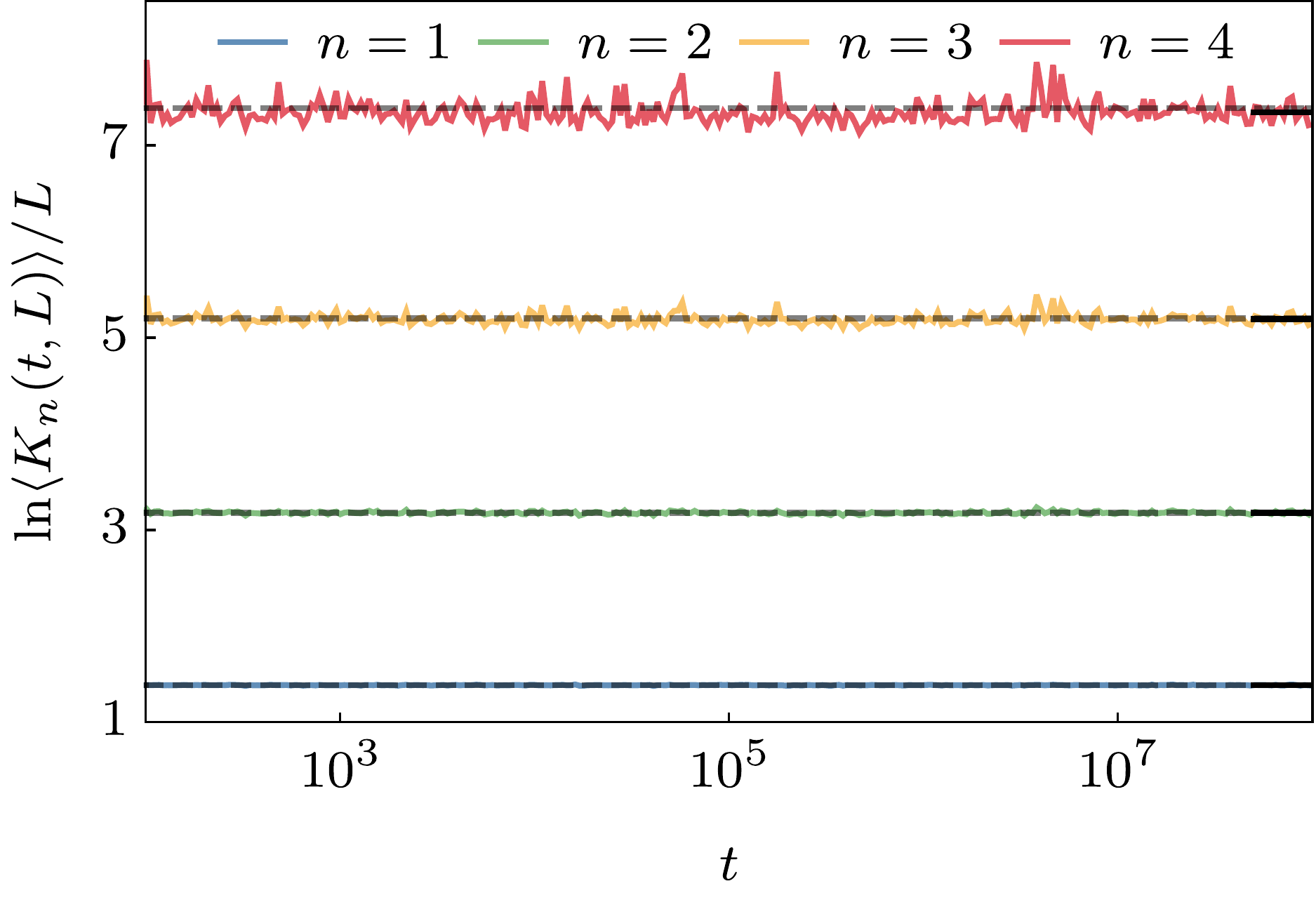}
   \caption{
Moments of the SFF versus time. With the dashed grey lines we indicate the predictions from Eq.~\eqref{eq:momentstinfty}, whereas black short lines indicate the average over time. We took $2L=10$, ${J_u=J_w=0.5}$, and averaged over $70000$ disorder realisations. }
    \label{fig:moments}
\end{figure}

Equation \eqref{eq:infinitetimeav} can also be used to find explicit predictions for higher moments of the spectral form factor at infinite times. Indeed, plugging \eqref{eq:infinitetimeav} into \eqref{eq:Tabcd} we find that ${\lim_{t\rightarrow \infty}\mathbb B_n}$ is unitarily equivalent to (cf. App.~\ref{sec:barBn})
\be
[\bar{\mathbb B}_n]_{a,b}:= \prod_{\mu,\nu=\pm}\!\!\! B(N_{\mu\nu}(a,b))\,,
\label{eq:Bntinfty}
\ee
where $N_{\mu\nu}(a,b)=\sum_{j=1}^{2n} \delta_{\mu,[a]_j}  \delta_{\nu,[b]_j}$. Diagonalising \eqref{eq:Bntinfty} numerically we then can access the infinite-time limit of generic moments \eqref{eq:SFFmoments}. Indeed
\be
\lim_{t\to\infty}K_n(t,L) = {\rm tr}[\bar{\mathbb B}_n^L] \simeq d_n \bar \lambda^L_{n}\,,
\label{eq:momentstinfty}
\ee
where we denoted by $\bar \lambda_{n}$ the largest magnitude eigenvalue of $\bar{\mathbb B}_n$ with multiplicity $d_n$ ($\simeq$ here represents the leading order in $L$). Explicit results for $\bar \lambda_{n}$ and $d_n$ for the first few values of $n$ are reported in Table~\ref{tab:lambdas}. The predictions of \eqref{eq:momentstinfty} are, once again, in agreement with our numerical experiments, see Fig.~\ref{fig:moments} for an example.

\section{Comparison with generic MBL systems at finite size}
\label{sec:MBL}

Interestingly, and in contrast with what is expected for generic localised systems, the result \eqref{eq:Bntinfty} is \emph{not} compatible with a Poissonian spectral statistics. Indeed, the moments of a Poisson random unitary matrix of size $2^{2L}$ read as ${K^{\rm P}_{n}(L) \simeq 2^{2 L n}} n! $~\cite{haake2018}, which coincides with \eqref{eq:momentstinfty} only for ${n=1}$. In particular, while $K^{\rm P}_{n}(L)/K^{\rm P}_{1}(L)^n$ is independent of $L$, $(\bar\lambda_{n}/\bar\lambda^n_{1})^L$ is exponentially large in $L$ signalling drastically enhanced fluctuations. This is the standard fingerprint of macroscopic collections of disconnected ``Poissonian patches'', namely of disconnected systems where each separate component is Poissonian~\footnote{Note that disconnected Poissonian systems do not follow the Poissonian statistics because the latter is not conserved under tensor multiplication. Too see this consider the tensor product of two large Poisson matrices of size $2^{L}$. This matrix has moments $K_n\simeq 2^{2 L n} (n!)^2$ which are not in Poisson form.}. For instance, taking a time evolution operator of the form 
\be
\mathbb U_{\xi,L}=\bigotimes_{x=1}^{2L/\xi} u^{(\xi)}_{x},
\label{eq:Upatch}
\ee
where $\{u^{(\xi)}_x\}$ are independent random diagonal unitary (i.e. Poisson random) matrices of size $2^\xi$, we find 
\be
K_{n,\xi}(L) =  \mathbb{E}\left[ |{\rm tr}[\mathbb U_{\xi,L}^t]|^{2n} \right]= r_n({\xi})^{2L}\,,
\label{eq:independentspins}
\ee
for any ${t\neq 0}$, where $r_n({1})=B(2n)$, $r_1({\xi})=2$, $r_n({\xi})>2^n$, and $r_n({\xi})<r_n({\xi-1})$ for ${n=2,\ldots,6}$~ (see App. \ref{sec:rn}).
 We then see that for large enough times (probing correlations among energy levels that are close) strongly localised circuits behave like collections of disconnected patches of size ${\xi>1}$. 

\begin{figure*}[ht!]
            (a)\!\!\!\!\raisebox{-\height+1em}{\includegraphics[width=.45\textwidth]{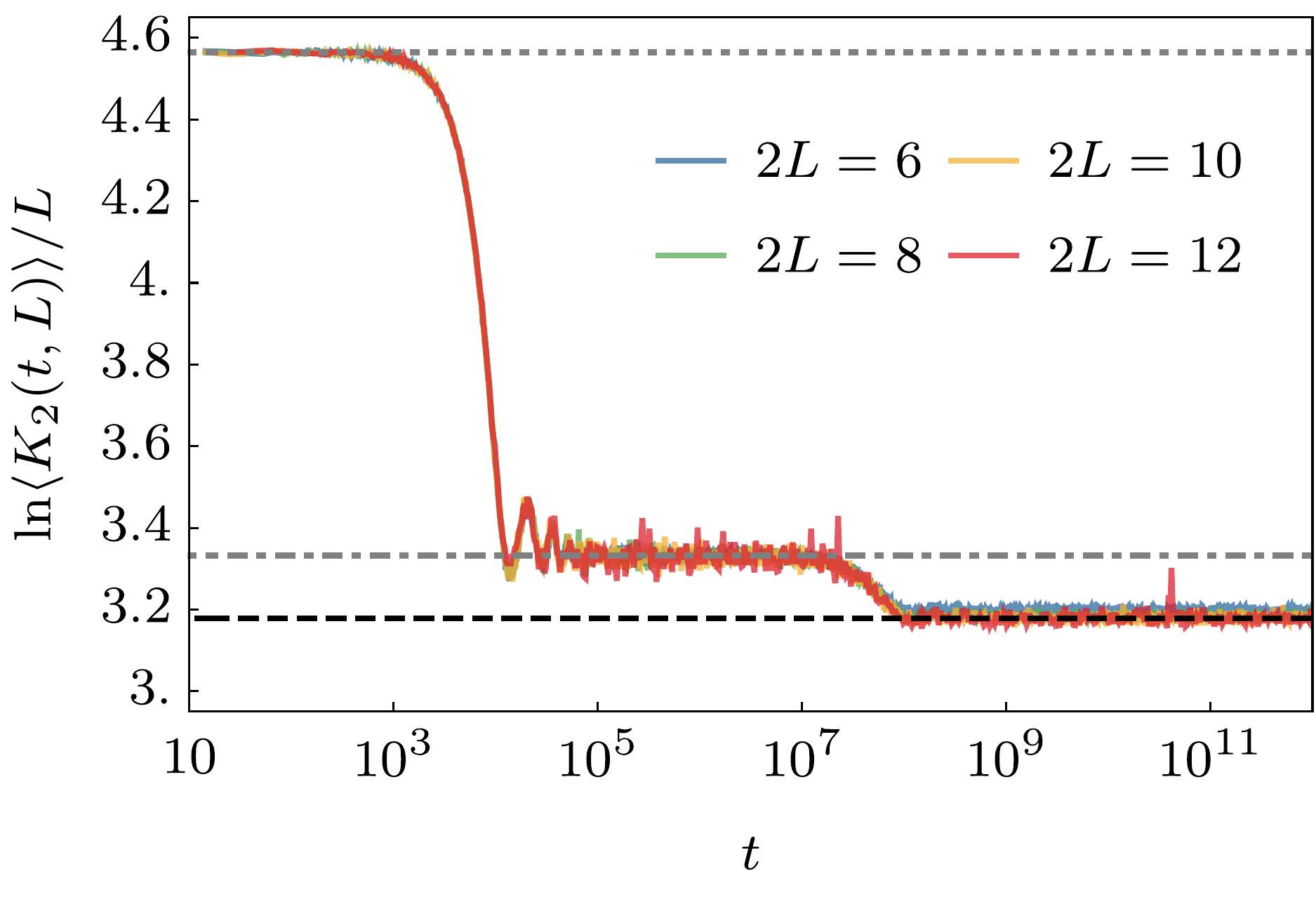}}%
           \hfill (b)\!\!\!\!\raisebox{-\height+.8em}{\includegraphics[width=.45\textwidth]{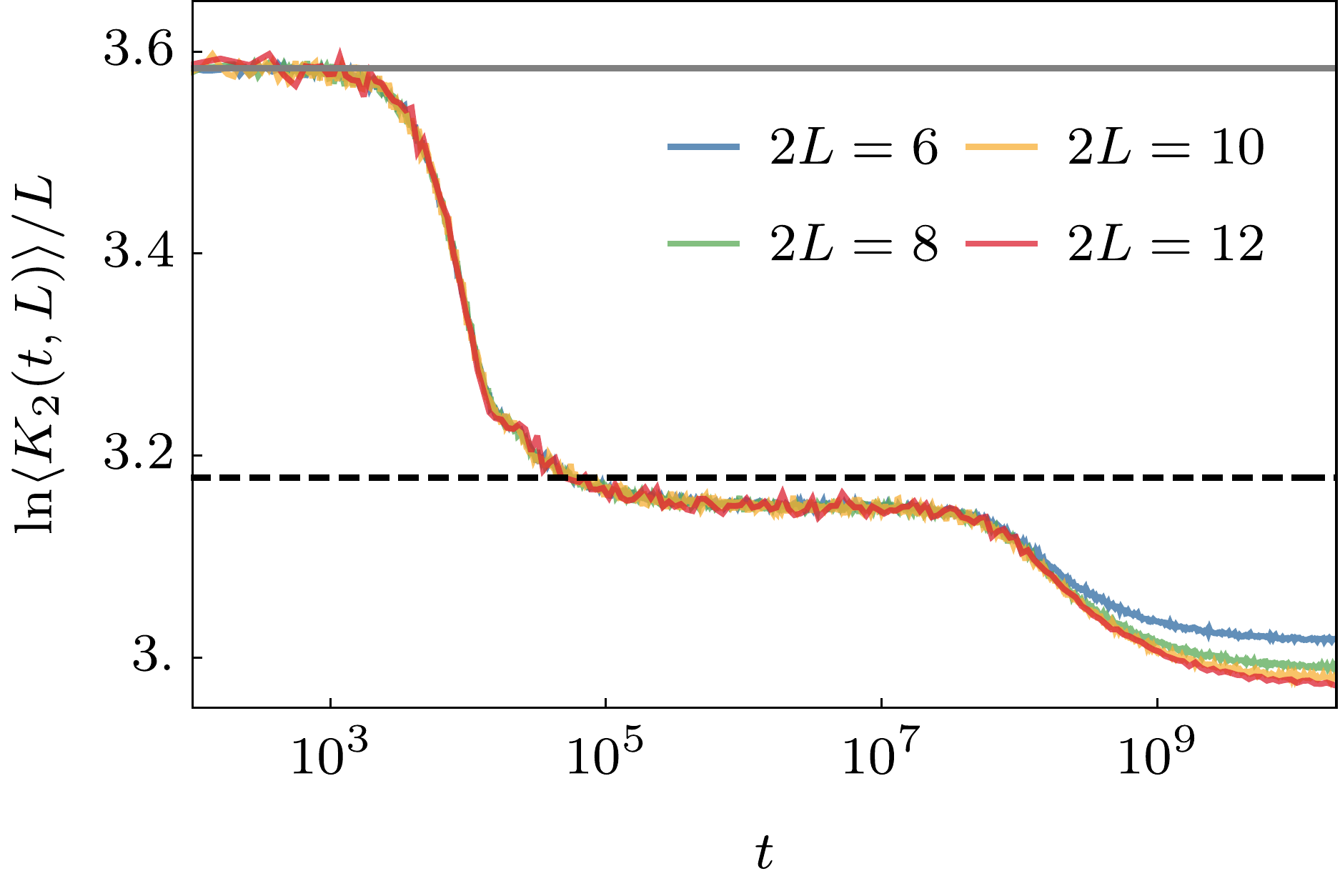}}
    \caption{Second moment of the SFF for strongly localised circuits (left) the circuit \eqref{eq:MBLcircuit} (right) versus time. Here we took ${J_u=J_w=J=10^{-4}}$, different coloured lines correspond to different system sizes. Dashed black, solid grey and dot-dashed grey lines indicate respectively the result \eqref{eq:momentstinfty}, \eqref{eq:independentspins} for $\xi=1$ and \eqref{eq:independentspins} for $\xi=2$. The dotted grey line shows $\log [B(2n) 4^n]$~\cite{Note4}. See \cite{Note5} for details about the averaging.}
    \label{fig:momentsMBL}
\end{figure*}

At larger energy separations (smaller times), however, the system looks like a collection of patches of \emph{smaller} size as demonstrated by the interesting cascade behaviour in Fig~\ref{fig:momentsMBL}a. In particular, we identify three different regimes: for times $t\ll \tau_1$ fluctuations match those of patches of size $\xi=1$~\footnote{In \eqref{eq:Upatch}, however, one has to take $\{u^{(\xi)}_{x}=\1\}_{x\in\mathbb Z}$. This agrees with \eqref{eq:Floquetgates} for $J_u=J_v=0$ and gives $K_{n,1}(L)= 4^n B(2n)$.}; for intermediate times ${\tau_1\ll t \ll \tau_2}$ they look like patches os size $2$; and for $t \gg \tau_2$ they approach the result \eqref{eq:momentstinfty}. For ${J_u=J_w=J}$ the crossover timescales are $\tau_1\propto J^{-1}$ and $\tau_2\propto J^{-2}$ (see App. \ref{sec:numerics}).

\footnotetext{ For the left panel we averaged over $10^5$ samples  ($10^4$ independently drawn disorder realisations and $10$ neighbouring time  samples for $2L=6,...,10$ and $3125$ independent samples ad $32$ time samples  for $2L=12$). For the right we averaged over $7\cdot 10^5$ realisations ($35\cdot 10^3$  independently drawn disorder realisations and $20$ neighbouring time samples  for $2L=6,...,10$ and $7\cdot 10^3$ independent samples and $10^2$ time samples for  $2L=12$}

It is natural to ask whether this is just a particular feature of strongly localised circuits or it characterises --- at least qualitatively --- all systems in the MBL regime. We studied this question numerically considering a ``generic" brickwork quantum circuit (cf.~\eqref{eq:Floquet}) with local gates of the form
\be
W_x=U_{x} = (u_{x-\frac{1}{2}} \otimes u_{x})\, U,
\label{eq:MBLcircuit}
\ee
where $U$ is given in Eq.~\eqref{eq:SL} and $u_x$ are Haar-random single-site unitary matrices. For fixed $L$ and small $J$ (i.e. strong disorder) this circuit exhibits MBL as signalled, e.g., by its Poissonian level spacing distribution (cf. App. \ref{sec:numerics})  (analogous MBL brickwork circuits have been studied in \cite{chan2018spectral, sunderhauf2018localization, garratt2021many, garratt2021local2}). The higher moments of the SFF, however, are not immediately Poissonian but show a cascade behaviour similar to that of strongly localised circuits, see e.g. Fig~\ref{fig:momentsMBL}b. Interestingly, as demonstrated by the scaling collapse in the figure, the first two plateaux are again compatible with \eqref{eq:independentspins}. In particular, the first one corresponds to patches of size $\xi=1$, while the second corresponds to an effective patch size $\xi_{\rm}\approx 3.45$. The crossover times scale with $J$ as in the strongly localised case.

We observed the emergence of a similar cascade behaviour (with the same collapse) also in other types of MBL systems at fixed $L$. In particular, in the Appendix \ref{sec:numerics} we present data for the disordered kicked Ising model (similar to the one studied in Ref.~\cite{FloquetMBL21}), a disordered spin chain with nearest-neighbour and next-nearest-neighbour Ising interaction, and a disordered $XYZ$ spin chain, which breaks $\mathbb{Z}_2$ symmetry.

Therefore, our analysis suggests that, in analogy with strongly localised circuits, generic MBL systems behave as collections of disconnected patches at intermediate energy scales. Considering correlations among sufficiently close energy levels, however, they show integrable-like (Poisson) spectral statistics. Note the value the size of the patches in the intermediate regime is model dependent. For instance, they are larger for next-nearest neighbour interactions (cf. App. \ref{sec:numerics}).

\section{Conclusions}
\label{sec:conclusions}

In summary, in this paper we presented a class of interacting localised many-body systems, dubbed ``strongly localised circuits", which are localised for any distribution of the disorder and whose spectral statistics can be characterised exactly. 
This parallels analogous exact results recently obtained for a special class of ergodic systems~\cite{PRL2018,CMP2021, flack2020statistics}. We showed that, even though the SFF of strongly localised circuits agrees with the Poissonian statistics, the higher moments are not Poissonian. We argued that this happens because these systems behave as macroscopic collections of disconnected Poissonian patches rather than a connected (Bethe-Ansatz) integrable system. Finally, we suggested that the same qualitative behaviour is observed in generic MBL systems at intermediate energy scales, while on shorter scales they display Poissonian spectral statistics.

The discovery of a class of solvable systems which are both interacting and localised paves the way for an exact analysis of localisation in the presence of interactions: Rigorous results in this controlled setting could fruitfully complement the numerous (mostly numerical) studies carried out over the last decade~\cite{nandkishore2015many, abanin2019many}.

A particularly interesting aspect is that the dynamics of strongly localised circuits are accessible via the time-channel description of Ref.~\cite{banulus2009matrix}. This opens the door to exact studies on the dynamics of local observables and quantum information, in analogy with recent results for special classes of integrable~\cite{pozsgay2013the, piroli2017what, piroli2018non, klobas2021exact, klobas2021exactII, klobas2021entanglement} and chaotic systems~\cite{gopalakrishnan2019unitary, bertini2019entanglement, piroli2020exact, claeys2021ergodic, lerose2020influence, suzuki2021computational}.

\section{Acknowledgments}
We thank Lorenzo Piroli for useful discussions and Adam Nahum for helpful comments on the manuscript. This work has been supported by the Royal Society through the University Research Fellowship No.\ 201101 (BB), European Research Council (ERC) through the Advanced Grant No.\ 694544 -- OMNES (PK and TP), and by Slovenian Research Agency (ARRS) through the Programme P1-0402 (TP). BB thanks the University of Ljubljana for hospitality during the completion of the project.

\onecolumngrid

\appendix

\section{Duality derivation of Eq.~(\ref{eq:SFFB})}
\label{sec:duality}

To evaluate \eqref{eq:SFFmoments} we adopt the diagrammatic representation introduced in \cite{CMP2021} and depict the SFF moments as  
\be
K_n(t,L)= \mathbb{E}\Biggl[
\begin{tikzpicture}[baseline=(current  bounding  box.center), scale=0.55]
\foreach \i in {2,...,5}{
\draw[very thick, dotted] (2*\i+2-12.5+0.255,-1.75-0.1) -- (2*\i+2-12.5+0.255,4.25-0.1);
\draw[very thick, dotted] (2*\i+2-11.5-0.255,-1.75-0.1) -- (2*\i+2-11.5-0.255,4.25-0.1);}

\foreach \i in {2,...,5}{
\draw[very thick] (2*\i+2-11.5,4) arc (-45:175:0.15);
\draw[very thick] (2*\i+2-11.5,-2) arc (315:180:0.15);
\draw[very thick] (2*\i+2-0.5-12,-2) arc (-135:0:0.15);
}
\foreach \i in {3,...,6}{
\draw[very thick] (2*\i+2-2.5-12,4) arc (225:0:0.15);
}
\foreach \i in {0,1,2}{
\draw[very thick, dotted] (-7.5,2*\i-1.745) -- (0.4,2*\i-1.745);
\draw[very thick, dotted] (-7.5,2*\i-1.255) -- (0.4,2*\i-1.255);
}
\foreach \i in{1.5,2.5,3.5}{
\draw[very thick] (0.5,2*\i-0.5-3.5) arc (45:-90:0.15);
\draw[very thick] (-8+0.5+0,2*\i-0.5-3.5) arc (45:270:0.15);
}
\foreach \i in{0.5,1.5,2.5}
{
\draw[very thick] (0.5,1+2*\i-0.5-3.5) arc (-45:90:0.15);
\draw[very thick] (-8+0.5,1+2*\i-0.5-3.5) arc (315:90:0.15);
}
\foreach \jj[evaluate=\jj as \j using -2*(ceil(\jj/2)-\jj/2)] in {-1,-3,-5}{
\foreach \i in {1,...,4}
{
\draw[very thick] (.5-2*\i-1*\j,-2-1*\jj) -- (1-2*\i-1*\j,-1.5-\jj);
\draw[very thick] (1-2*\i-1*\j,-1.5-1*\jj) -- (1.5-2*\i-1*\j,-2-\jj);
\draw[very thick] (.5-2*\i-1*\j,-1-1*\jj) -- (1-2*\i-1*\j,-1.5-\jj);
\draw[very thick] (1-2*\i-1*\j,-1.5-1*\jj) -- (1.5-2*\i-1*\j,-1-\jj);
\draw[thick, fill=OliveGreen, rounded corners=2pt] (0.75-2*\i-1*\j,-1.75-\jj) rectangle (1.25-2*\i-1*\j,-1.25-\jj);
\draw[ thick] (-2*\i+2,-1.35-\jj) -- (-2*\i+2.15,-1.35-\jj) -- (-2*\i+2.15,-1.5-\jj);%
}
}
\foreach \jj[evaluate=\jj as \j using -2*(ceil(\jj/2)-\jj/2)] in {-4,-2,0}{
\foreach \i in {1,...,4}
{
\draw[very thick] (.5-2*\i-1*\j,-2-1*\jj) -- (1-2*\i-1*\j,-1.5-\jj);
\draw[very thick] (1-2*\i-1*\j,-1.5-1*\jj) -- (1.5-2*\i-1*\j,-2-\jj);
\draw[very thick] (.5-2*\i-1*\j,-1-1*\jj) -- (1-2*\i-1*\j,-1.5-\jj);
\draw[very thick] (1-2*\i-1*\j,-1.5-1*\jj) -- (1.5-2*\i-1*\j,-1-\jj);
\draw[thick, fill=mygreen, rounded corners=2pt] (0.75-2*\i-1*\j,-1.75-\jj) rectangle (1.25-2*\i-1*\j,-1.25-\jj);
\draw[ thick] (-2*\i+1,-1.35-\jj) -- (-2*\i+1.15,-1.35-\jj) -- (-2*\i+1.15,-1.5-\jj);%
}
}
\foreach \jj in {0,2,4}{
\draw[ thick, fill=myyellow1, rounded corners=2pt] (0.5,-1+\jj) circle (.15);
\draw[ thick, fill=myyellow3, rounded corners=2pt] (-1.5,-1+\jj) circle (.15);
\draw[ thick, fill=myyellow5, rounded corners=2pt] (-3.5,-1+\jj) circle (.15);
\draw[ thick, fill=myyellow7, rounded corners=2pt] (-5.5,-1+\jj) circle (.15);
\draw[ thick, fill=myyellow8, rounded corners=2pt] (0.5,\jj) circle (.15);
\draw[ thick, fill=myyellow1, rounded corners=2pt] (-1.5,\jj) circle (.15);
\draw[ thick, fill=myyellow2, rounded corners=2pt] (-3.5,\jj) circle (.15);
\draw[ thick, fill=myyellow10, rounded corners=2pt] (-5.5,\jj) circle (.15);
}
\foreach \jj in {0,-2,-4}{
\foreach \i in {1,...,4}{
\draw[ thick] (-2*\i+2.4,0.05-\jj) -- (-2*\i+2.55,0.05-\jj) -- (-2*\i+2.55,-0.1-\jj);
}
\foreach \i in {0,...,3}{
\draw[ thick]  (-2*\i+.45,-1.1-\jj) -- (-2*\i+.45,-0.95-\jj) -- (-2*\i+.6,-0.95-\jj);
}
}
\end{tikzpicture}\Biggr]\,,
\label{eq:SFFfolded}
\ee
where we introduced $2n$-folded gates
\begin{subequations}
\begin{align}
\label{eq:doublegate}
&\begin{tikzpicture}[baseline=(current  bounding  box.center), scale=.7]
\def\eps{0.5};
\Wgategreen{-3.75}{0};
\Text[x=-3.75,y=-0.6]{}
\end{tikzpicture}
= U^{\otimes n}\!\otimes\! (U^{*})^{\otimes n}\!,
\quad
\begin{tikzpicture}[baseline=(current  bounding  box.center), scale=.7]
\def\eps{0.5};
\Wgateolivegreen{-3.75}{0};
\Text[x=-3.75,y=-0.6]{}
\end{tikzpicture}
= W^{\otimes n}\!\otimes\! (W^{*})^{\otimes n}\!, \\ 
&
\begin{tikzpicture}[baseline=(current  bounding  box.center), scale=.7]
\draw[very thick] (-4.25,0.5) -- (-4.25,-0.5);
\draw[ thick, fill=myYO, rounded corners=2pt] (-4.25,0) circle (.15);
\draw[thick, rotate around = {-45:(0.525-4.77,0.375-0.4)}]  (.45-4.77,0.3-0.4) -- (.45-4.77,0.45-0.4) -- (.6-4.77,0.45-0.4);
\Text[x=-4.25,y=-0.75]{}
\end{tikzpicture}
= (u_{x})^{\otimes n}\!\otimes\! (u_{x}^{*})^{\otimes n},\,\, (u_{x})^{\otimes n}\!\otimes\! (u_{x}^{*})^{\otimes n}\,.
\end{align}
\end{subequations}
Here and in the following $(\cdot)^*$ denotes complex conjugation in the canonical basis. Using that  $u_{x}, w_{x}$ are independently distributed in space we have that $K_n(t,L)$ can be written in terms of a spatial transfer matrix, i.e.
\begin{align}
\!K_n(t,L) & \!=\! {\rm tr}\, \mathbb{E}\Bigl[\begin{tikzpicture}[baseline=(current  bounding  box.center), scale=0.5]
\foreach \i in {5}{
\draw[very thick, dotted] (2*\i+2-12.5+0.255,-1.75-0.1) -- (2*\i+2-12.5+0.255,4.25-0.1);
\draw[very thick, dotted] (2*\i+2-11.5-0.255,-1.75-0.1) -- (2*\i+2-11.5-0.255,4.25-0.1);}
\foreach \i in {5}{
\draw[very thick] (2*\i+2-11.5,4) arc (-45:175:0.15);
\draw[very thick] (2*\i+2-11.5,-2) arc (315:180:0.15);
\draw[very thick] (2*\i+2-0.5-12,-2) arc (-135:0:0.15);
}
\foreach \i in {6}{
\draw[very thick] (2*\i+2-2.5-12,4) arc (225:0:0.15);
}
\foreach \jj[evaluate=\jj as \j using -2*(ceil(\jj/2)-\jj/2)] in {-1,-3,-5}{
\foreach \i in {1}
{
\draw[very thick] (.5-2*\i-1*\j,-2-1*\jj) -- (1-2*\i-1*\j,-1.5-\jj);
\draw[very thick] (1-2*\i-1*\j,-1.5-1*\jj) -- (1.5-2*\i-1*\j,-2-\jj);
\draw[very thick] (.5-2*\i-1*\j,-1-1*\jj) -- (1-2*\i-1*\j,-1.5-\jj);
\draw[very thick] (1-2*\i-1*\j,-1.5-1*\jj) -- (1.5-2*\i-1*\j,-1-\jj);
\draw[thick, fill=OliveGreen, rounded corners=2pt] (0.75-2*\i-1*\j,-1.75-\jj) rectangle (1.25-2*\i-1*\j,-1.25-\jj);
\draw[ thick] (-2*\i+2,-1.35-\jj) -- (-2*\i+2.15,-1.35-\jj) -- (-2*\i+2.15,-1.5-\jj);%
}
}
\foreach \jj[evaluate=\jj as \j using -2*(ceil(\jj/2)-\jj/2)] in {-4,-2,0}{
\foreach \i in {1}
{
\draw[very thick] (.5-2*\i-1*\j,-2-1*\jj) -- (1-2*\i-1*\j,-1.5-\jj);
\draw[very thick] (1-2*\i-1*\j,-1.5-1*\jj) -- (1.5-2*\i-1*\j,-2-\jj);
\draw[very thick] (.5-2*\i-1*\j,-1-1*\jj) -- (1-2*\i-1*\j,-1.5-\jj);
\draw[very thick] (1-2*\i-1*\j,-1.5-1*\jj) -- (1.5-2*\i-1*\j,-1-\jj);
\draw[thick, fill=mygreen, rounded corners=2pt] (0.75-2*\i-1*\j,-1.75-\jj) rectangle (1.25-2*\i-1*\j,-1.25-\jj);
\draw[ thick] (-2*\i+1,-1.35-\jj) -- (-2*\i+1.15,-1.35-\jj) -- (-2*\i+1.15,-1.5-\jj);%
}
}
\foreach \jj in {0,2,4}{
\draw[ thick, fill=myyellow1, rounded corners=2pt] (0.5,-1+\jj) circle (.15);
\draw[ thick, fill=myyellow8, rounded corners=2pt] (0.5,\jj) circle (.15);
}
\foreach \jj in {0,-2,-4}{
\foreach \i in {1}{
}
\foreach \i in {1}{
\draw[ thick] (-2*\i+2.4,0.05-\jj) -- (-2*\i+2.55,0.05-\jj) -- (-2*\i+2.55,-0.1-\jj);}
\foreach \i in {1}{
}
\foreach \i in {0}{
\draw[ thick]  (-2*\i+.45,-1.1-\jj) -- (-2*\i+.45,-0.95-\jj) -- (-2*\i+.6,-0.95-\jj);}
}
\end{tikzpicture}\Bigr]\mathbb{E}\Bigl[\begin{tikzpicture}[baseline=(current  bounding  box.center), scale=0.5]
\foreach \i in {5}{
\draw[very thick, dotted] (2*\i+2-12.5+0.255,-1.75-0.1) -- (2*\i+2-12.5+0.255,4.25-0.1);
\draw[very thick, dotted] (2*\i+2-11.5-0.255,-1.75-0.1) -- (2*\i+2-11.5-0.255,4.25-0.1);}
\foreach \i in {5}{
\draw[very thick] (2*\i+2-11.5,4) arc (-45:175:0.15);
\draw[very thick] (2*\i+2-11.5,-2) arc (315:180:0.15);
\draw[very thick] (2*\i+2-0.5-12,-2) arc (-135:0:0.15);
}
\foreach \i in {6}{
\draw[very thick] (2*\i+2-2.5-12,4) arc (225:0:0.15);
}
\foreach \jj[evaluate=\jj as \j using -2*(ceil(\jj/2)-\jj/2)] in {-1,-3,-5}{
\foreach \i in {1}
{
\draw[very thick] (.5-2*\i-1*\j,-2-1*\jj) -- (1-2*\i-1*\j,-1.5-\jj);
\draw[very thick] (1-2*\i-1*\j,-1.5-1*\jj) -- (1.5-2*\i-1*\j,-2-\jj);
\draw[very thick] (.5-2*\i-1*\j,-1-1*\jj) -- (1-2*\i-1*\j,-1.5-\jj);
\draw[very thick] (1-2*\i-1*\j,-1.5-1*\jj) -- (1.5-2*\i-1*\j,-1-\jj);
\draw[thick, fill=OliveGreen, rounded corners=2pt] (0.75-2*\i-1*\j,-1.75-\jj) rectangle (1.25-2*\i-1*\j,-1.25-\jj);
\draw[ thick] (-2*\i+2,-1.35-\jj) -- (-2*\i+2.15,-1.35-\jj) -- (-2*\i+2.15,-1.5-\jj);%
}
}
\foreach \jj[evaluate=\jj as \j using -2*(ceil(\jj/2)-\jj/2)] in {-4,-2,0}{
\foreach \i in {1}
{
\draw[very thick] (.5-2*\i-1*\j,-2-1*\jj) -- (1-2*\i-1*\j,-1.5-\jj);
\draw[very thick] (1-2*\i-1*\j,-1.5-1*\jj) -- (1.5-2*\i-1*\j,-2-\jj);
\draw[very thick] (.5-2*\i-1*\j,-1-1*\jj) -- (1-2*\i-1*\j,-1.5-\jj);
\draw[very thick] (1-2*\i-1*\j,-1.5-1*\jj) -- (1.5-2*\i-1*\j,-1-\jj);
\draw[thick, fill=mygreen, rounded corners=2pt] (0.75-2*\i-1*\j,-1.75-\jj) rectangle (1.25-2*\i-1*\j,-1.25-\jj);
\draw[ thick] (-2*\i+1,-1.35-\jj) -- (-2*\i+1.15,-1.35-\jj) -- (-2*\i+1.15,-1.5-\jj);%
}
}
\foreach \jj in {0,2,4}{
\draw[ thick, fill=myyellow1, rounded corners=2pt] (0.5,-1+\jj) circle (.15);
\draw[ thick, fill=myyellow8, rounded corners=2pt] (0.5,\jj) circle (.15);
}
\foreach \jj in {0,-2,-4}{
\foreach \i in {1}{
}
\foreach \i in {1}{
\draw[ thick] (-2*\i+2.4,0.05-\jj) -- (-2*\i+2.55,0.05-\jj) -- (-2*\i+2.55,-0.1-\jj);}
\foreach \i in {1}{
}
\foreach \i in {0}{
\draw[ thick]  (-2*\i+.45,-1.1-\jj) -- (-2*\i+.45,-0.95-\jj) -- (-2*\i+.6,-0.95-\jj);}
}
\end{tikzpicture}\Bigr]\mathbb{E}\Bigl[\begin{tikzpicture}[baseline=(current  bounding  box.center), scale=0.5]
\foreach \i in {5}{
\draw[very thick, dotted] (2*\i+2-12.5+0.255,-1.75-0.1) -- (2*\i+2-12.5+0.255,4.25-0.1);
\draw[very thick, dotted] (2*\i+2-11.5-0.255,-1.75-0.1) -- (2*\i+2-11.5-0.255,4.25-0.1);}
\foreach \i in {5}{
\draw[very thick] (2*\i+2-11.5,4) arc (-45:175:0.15);
\draw[very thick] (2*\i+2-11.5,-2) arc (315:180:0.15);
\draw[very thick] (2*\i+2-0.5-12,-2) arc (-135:0:0.15);
}
\foreach \i in {6}{
\draw[very thick] (2*\i+2-2.5-12,4) arc (225:0:0.15);
}
\foreach \jj[evaluate=\jj as \j using -2*(ceil(\jj/2)-\jj/2)] in {-1,-3,-5}{
\foreach \i in {1}
{
\draw[very thick] (.5-2*\i-1*\j,-2-1*\jj) -- (1-2*\i-1*\j,-1.5-\jj);
\draw[very thick] (1-2*\i-1*\j,-1.5-1*\jj) -- (1.5-2*\i-1*\j,-2-\jj);
\draw[very thick] (.5-2*\i-1*\j,-1-1*\jj) -- (1-2*\i-1*\j,-1.5-\jj);
\draw[very thick] (1-2*\i-1*\j,-1.5-1*\jj) -- (1.5-2*\i-1*\j,-1-\jj);
\draw[thick, fill=OliveGreen, rounded corners=2pt] (0.75-2*\i-1*\j,-1.75-\jj) rectangle (1.25-2*\i-1*\j,-1.25-\jj);
\draw[ thick] (-2*\i+2,-1.35-\jj) -- (-2*\i+2.15,-1.35-\jj) -- (-2*\i+2.15,-1.5-\jj);%
}
}
\foreach \jj[evaluate=\jj as \j using -2*(ceil(\jj/2)-\jj/2)] in {-4,-2,0}{
\foreach \i in {1}
{
\draw[very thick] (.5-2*\i-1*\j,-2-1*\jj) -- (1-2*\i-1*\j,-1.5-\jj);
\draw[very thick] (1-2*\i-1*\j,-1.5-1*\jj) -- (1.5-2*\i-1*\j,-2-\jj);
\draw[very thick] (.5-2*\i-1*\j,-1-1*\jj) -- (1-2*\i-1*\j,-1.5-\jj);
\draw[very thick] (1-2*\i-1*\j,-1.5-1*\jj) -- (1.5-2*\i-1*\j,-1-\jj);
\draw[thick, fill=mygreen, rounded corners=2pt] (0.75-2*\i-1*\j,-1.75-\jj) rectangle (1.25-2*\i-1*\j,-1.25-\jj);
\draw[ thick] (-2*\i+1,-1.35-\jj) -- (-2*\i+1.15,-1.35-\jj) -- (-2*\i+1.15,-1.5-\jj);%
}
}
\foreach \jj in {0,2,4}{
\draw[ thick, fill=myyellow1, rounded corners=2pt] (0.5,-1+\jj) circle (.15);
\draw[ thick, fill=myyellow8, rounded corners=2pt] (0.5,\jj) circle (.15);
}
\foreach \jj in {0,-2,-4}{
\foreach \i in {1}{
}
\foreach \i in {1}{
\draw[ thick] (-2*\i+2.4,0.05-\jj) -- (-2*\i+2.55,0.05-\jj) -- (-2*\i+2.55,-0.1-\jj);}
\foreach \i in {1}{
}
\foreach \i in {0}{
\draw[ thick]  (-2*\i+.45,-1.1-\jj) -- (-2*\i+.45,-0.95-\jj) -- (-2*\i+.6,-0.95-\jj);}
}
\end{tikzpicture}\Bigr]\mathbb{E}\Bigl[\begin{tikzpicture}[baseline=(current  bounding  box.center), scale=0.5]
\foreach \i in {5}{
\draw[very thick, dotted] (2*\i+2-12.5+0.255,-1.75-0.1) -- (2*\i+2-12.5+0.255,4.25-0.1);
\draw[very thick, dotted] (2*\i+2-11.5-0.255,-1.75-0.1) -- (2*\i+2-11.5-0.255,4.25-0.1);}
\foreach \i in {5}{
\draw[very thick] (2*\i+2-11.5,4) arc (-45:175:0.15);
\draw[very thick] (2*\i+2-11.5,-2) arc (315:180:0.15);
\draw[very thick] (2*\i+2-0.5-12,-2) arc (-135:0:0.15);
}
\foreach \i in {6}{
\draw[very thick] (2*\i+2-2.5-12,4) arc (225:0:0.15);
}
\foreach \jj[evaluate=\jj as \j using -2*(ceil(\jj/2)-\jj/2)] in {-1,-3,-5}{
\foreach \i in {1}
{
\draw[very thick] (.5-2*\i-1*\j,-2-1*\jj) -- (1-2*\i-1*\j,-1.5-\jj);
\draw[very thick] (1-2*\i-1*\j,-1.5-1*\jj) -- (1.5-2*\i-1*\j,-2-\jj);
\draw[very thick] (.5-2*\i-1*\j,-1-1*\jj) -- (1-2*\i-1*\j,-1.5-\jj);
\draw[very thick] (1-2*\i-1*\j,-1.5-1*\jj) -- (1.5-2*\i-1*\j,-1-\jj);
\draw[thick, fill=OliveGreen, rounded corners=2pt] (0.75-2*\i-1*\j,-1.75-\jj) rectangle (1.25-2*\i-1*\j,-1.25-\jj);
\draw[ thick] (-2*\i+2,-1.35-\jj) -- (-2*\i+2.15,-1.35-\jj) -- (-2*\i+2.15,-1.5-\jj);%
}
}
\foreach \jj[evaluate=\jj as \j using -2*(ceil(\jj/2)-\jj/2)] in {-4,-2,0}{
\foreach \i in {1}
{
\draw[very thick] (.5-2*\i-1*\j,-2-1*\jj) -- (1-2*\i-1*\j,-1.5-\jj);
\draw[very thick] (1-2*\i-1*\j,-1.5-1*\jj) -- (1.5-2*\i-1*\j,-2-\jj);
\draw[very thick] (.5-2*\i-1*\j,-1-1*\jj) -- (1-2*\i-1*\j,-1.5-\jj);
\draw[very thick] (1-2*\i-1*\j,-1.5-1*\jj) -- (1.5-2*\i-1*\j,-1-\jj);
\draw[thick, fill=mygreen, rounded corners=2pt] (0.75-2*\i-1*\j,-1.75-\jj) rectangle (1.25-2*\i-1*\j,-1.25-\jj);
\draw[ thick] (-2*\i+1,-1.35-\jj) -- (-2*\i+1.15,-1.35-\jj) -- (-2*\i+1.15,-1.5-\jj);%
}
}
\foreach \jj in {0,2,4}{
\draw[ thick, fill=myyellow1, rounded corners=2pt] (0.5,-1+\jj) circle (.15);
\draw[ thick, fill=myyellow8, rounded corners=2pt] (0.5,\jj) circle (.15);
}
\foreach \jj in {0,-2,-4}{
\foreach \i in {1}{
}
\foreach \i in {1}{
\draw[ thick] (-2*\i+2.4,0.05-\jj) -- (-2*\i+2.55,0.05-\jj) -- (-2*\i+2.55,-0.1-\jj);}
\foreach \i in {1}{
}
\foreach \i in {0}{
\draw[ thick]  (-2*\i+.45,-1.1-\jj) -- (-2*\i+.45,-0.95-\jj) -- (-2*\i+.6,-0.95-\jj);}
}
\end{tikzpicture}\Bigr] \notag\\
&= {\rm tr}\,\mathbb T_n^L.
 \label{eq:SFFduality}
\end{align}
In formulas, the transfer matrix can be expressed as 
\be
\mathbb T_n= [\tilde{\mathbb U}^{\otimes n} \!\otimes \tilde{\mathbb U}^{\ast \otimes n}]  [\tilde{\mathbb W}^{\otimes n} \!\otimes \tilde{\mathbb W}^{\ast \otimes n}]\mathbb O_n^{\phantom{\dag}},
\ee
where we introduced the non-expanding operator 
\be
\mathbb O_n = \mathbb{E}\!\!\left[ \prod_{\tau \in \mathbb Z_{t}+\frac{1}{2}} \!\eta_{\tau,t}(u\otimes w)^{\otimes n}\otimes \eta_{\tau,t}(u^*\otimes w^*)^{\otimes n}\right]\!\!
\label{eq:defOa}
\ee
and the unitary operators 
\be
\tilde{\mathbb U} := \!\!\prod_{\tau \in \mathbb Z_{t}+\frac{1}{2}}\!\!\eta_{\tau,t}(\tilde U),\quad
\tilde{\mathbb W} := \prod_{\tau \in \mathbb Z_{t}} \eta_{\tau,t}(\tilde W),
\ee
constructed in terms of the ``dual" local gates $\tilde U, \tilde W$. The latter are obtained applying the reshuffling transformation $\tilde{(\cdot)}$, such that 
\be
[\tilde O]_{ij,kl}= O_{ki,lj},\qquad i,j,k,l=1,2\,,
\ee
to the local gates $U, W$. Note that in the above equations we used the positioning operator on a chain of length $2t$ 
\be
\eta_{x,t}(O) = \Pi_{2t}^{2x-|O|+1} (O \otimes\1_{2t-|O|}) \Pi_{2t}^{-2x+|O|-1}
\ee
where $|O|$ is the support size of $O$, while $\1_{\ell}$ and $\Pi_\ell$ are respectively the identity and the operator implementing a periodic one-site shift to the right in a chain of $\ell$ qubits.

The explicit form \eqref{eq:SL} yields
\be
\tilde U = \!\sum_{s\in\{0,1\}}\!C_{us} \ket{Z^s}\bra{Z^s}, 
\qquad\qquad
\tilde W =\!\sum_{s\in\{0,1\}}\!C_{ws} \ket{Z^s}\bra{Z^s},
\ee
where $C_{\iota s} \!\!= i^s \cos(J_\iota+ s \tfrac{\pi}{2})$, $\iota=u,w$, and the two-site states are vectorized matrices
\be
\bra{A} = \begin{tikzpicture}[baseline={([yshift=-.5ex]current bounding box.center)}, scale=0.45]
\draw[thick] (0.65,0.2) arc (90:270:0.45cm);  
\draw[thick, fill=black] (0.2,-0.25) circle (0.05cm);
\Text[x=-0.2,y=-0.2]{$A$}
\end{tikzpicture}\,, \qquad A=I,X,Y,Z\,.
\ee
This means that the many-body operators $\tilde{\mathbb U}, \tilde{\mathbb W}$ can be written as 
\begin{align}
&\tilde{\mathbb U} =  \sum_{s_j\in\{0,1\}} \!\!\!C_{u s_1}\!\cdots C_{u s_t} \bigotimes_{j=1}^t \ket{Z^{s_j}}\!\!\bra{Z^{s_j}},
&
&\tilde{\mathbb W} = \sum_{s_j \in\{0,1\}} \!\!\!C_{w s_1}\!\cdots C_{w s_t} \Pi_{2t}\!\!\left[\bigotimes_{j=1}^t \ket{Z^{s_j}}\!\!\bra{Z^{s_j}}\right]\!\! \Pi^\dag_{2t}\,.  
\end{align}
Now, noting that  
\begin{align}
\bra{Z^{s_1},Z^{s_2},\cdots Z^{s_t}} \Pi_{2t} \ket{Z^{r_1},Z^{r_2},\cdots Z^{r_t}} = {\rm tr}[Z^{s_1}Z^{r_1}\cdots Z^{s_t}Z^{r_t}]\! = \!1+(-1)^{\sum_{j=1}^t (r_j+s_j)},
\end{align}  
 we find 
\be
\tilde{\mathbb U} \tilde{\mathbb W} = \ket{\Phi_{u0}}\bra{\Phi_{w0}} \Pi^\dag_{2t}+ \ket{\Phi_{u1}}\bra{\Phi_{w1}} \Pi^\dag_{2t}, 
\label{eq:UW}
\ee
where we introduced the un-normalised orthogonal vectors 
\begin{align}
\ket{\Phi_{u a}} & := \sqrt{2}\!\!\!\!\! \sum_{\substack{s_1,...., s_t\in\{0,1\} \\ s_1+\ldots + s_t = a\, {\rm mod}\,2}} \!\!\!\!\!\!\! C_{u s_1}\cdots C_{u s_t} \ket{Z^{s_1}\cdots Z^{s_t}}\!,\\
\ket{\Phi_{w a}} & := \sqrt{2}\!\!\!\!\! \sum_{\substack{s_1,...., s_t\in\{0,1\} \\ s_1+\ldots + s_t = a\, {\rm mod}\,2}} \!\!\!\!\!\!\! C^{\ast}_{w s_1}\cdots C^{\ast}_{w s_t} \ket{Z^{s_1}\cdots Z^{s_t}}\!.
\end{align}
The form \eqref{eq:UW} is the key for our analytical derivation as it shows that $\tilde{\mathbb U} \tilde{\mathbb W}$ is a projector on a 2-dimensional subspace. Indeed, this means that $\mathbb T_n$ is the tensor product of $2n$ such projectors and can have at most $4^n$ non-trivial eigenvalues. In particular, recalling the definition \eqref{eq:defOa} of $\mathbb O_n$ we have that the non-trivial $4^n\times4^n$ block $\mathbb B_n$ of $\mathbb T_n$ is written as 
\be
[\mathbb B_n]_{a,b} =\mathbb{E}\!\left[ \prod_{j=1}^n\! M^{\phantom{\ast}}_{[a]_j [b]_j} M^*_{[a]_{n+j} [b]_{n+j}}\!\right]\!\!,
\label{eq:Tabcd2}
\ee
where $a,b=0,\ldots,4^n-1$, $[x]_j$ is the $j$-th digit of $x$ in base 2, and we introduced 
\be
M_{ab}:=\braket{\Phi_{wa}|\Pi^\dag_{2t} \!\!\prod_{\tau \in \mathbb Z_{t}+\frac{1}{2}}\!\!\!\!\eta_{\tau,t}(u\otimes w) |\Phi_{ub}}. 
\label{eq:Mab}
\ee
To conclude we evaluate this expression using the explicit form of the states $\ket{\Phi_{u a}}$, $\ket{\Phi_{w a}}$. We start by expressing it explicitly as follows 
\be
M_{ab} = 2  \sum_{\substack{s_1,...., s_t\in\{0,1\} \\ s_1+\ldots + s_t = a\, {\rm mod}\, 2}} \sum_{\substack{r_1,...., r_t\in\{0,1\} \\ r_1+\ldots + r_t = b\, {\rm mod}\, 2}} \!\!\!\!\!  C_{w s_1}\cdots C_{w s_t}C_{u r_1}\cdots C_{u r_t} {\rm tr}[Z^{s_1} u Z^{r_1} w Z^{s_2} u Z^{r_2}\cdots Z^{s_t} u Z^{r_t} w]. 
\ee
The constraints can be treated by facilitating a simple identity
\be
\frac{1}{2}\sum_{\mu=\pm} \mu^K = \delta_{K\,{\rm mod}\, 2,0}\,,
\ee
which gives 
\begin{align}
M_{ab} =& \frac{1}{2} \sum_{\mu_j=\pm}\sum_{{s_j\in\{0,1\}}} \sum_{\substack{r_j\in\{0,1\}}} \mu_1^{s_1+\ldots + s_t+a} \mu_2^{r_1+\ldots + r_t+b} C_{w s_1}\cdots C_{w s_t}C_{u r_1}\cdots C_{u r_t}  {\rm tr}[Z^{s_1} u Z^{r_1} w Z^{s_2} u Z^{r_2}\cdots Z^{s_t} u Z^{r_t} w]\notag\\
=& \frac{1}{2} \sum_{\mu,\nu =\pm}\!\! \mu^a \nu^b\, {\rm tr}[(w e^{i \mu J_w Z} u e^{i \nu J_u Z} )^t] =  \sum_{\mu,\nu =\pm}\!\! \mu^a \nu^b\! \cos(\phi_{\mu\nu}t)\,.
\label{eq:Mabsimp2}
\end{align}
In the second step we denoted by $\exp[\pm i \phi_{\mu\nu}]$ the eigenvalues of \eqref{eq:vdef}. 

This matrix can be considered as element of ${\rm SU}(2)$ since one may choose $u,v\in {\rm SU}(2)$ (a global phase multiplying these matrices can be absorbed in \eqref{eq:Tabcd}). In particular, considering the Euler-angle representation of $u$ and $w$
\be
u = u(a_1,b_1,c_1) \equiv e^{i a_1 Z/2} e^{i b_1 Y/2} e^{i c_1 Z/2},\qquad w = u(a_2,b_2,c_2),
\label{eq:Eulervw}
\ee
with $a_j\in[0,2\pi]$, $b_j\in[0,\pi]$, and $c_j\in[0,4\pi]$ we have 
\be
\!\!\!\phi_{\eta\nu} = {\rm atan}\sqrt{\frac{1}{\Delta^2}-1}+ \pi (1-\theta(\Delta))
\ee
where $\theta(x)$ is the Heaviside function and we set 
\begin{align}
\Delta =& \cos\left(\frac{b_1}{2}\right) \cos\left(\frac{b_2}{2}\right)\cos\left(\frac{a_1+a_2+c_1+c_2+2\eta J_u +2\nu J_w}{2}\right)\notag\\
&- \sin\left(\frac{b_1}{2}\right) \sin\left(\frac{b_2}{2}\right)\cos\left(\frac{a_1-a_2-c_1+c_2-2\eta J_u +2\nu J_w}{2}\right)\,.
\end{align}

\section{Explicit form of $\mathbb B_1$ for Haar random $u_x, w_x$}
\label{sec:B1haar}

Let us assume that $\mathbb{E}[\cdot]$ is the average over the Haar measure of ${\rm SU}(2)$. In this case, using the left and right translation invariance of the measure we find 
\be
\mathbb{E}[{\rm tr}[(u e^{i \mu_1 J_u Z} w e^{i \nu_1 J_w Z})^t]({\rm tr}[(u e^{i \mu_2 J_u Z} w e^{i \nu_2 J_w Z})^t])^*]= \mathbb{E}[{\rm tr}[(u w)^t]({\rm tr}[(u e^{i (\mu_2-\mu_1) J_u Z} w e^{i (\nu_2-\nu_1) J_w Z})^t])^*].
\ee
Considering the cases $\mu_2=\pm\mu_1$ and $\nu_2=\pm\nu_1$ separately we have 
\begin{align}
\mathbb{E}[{\rm tr}[(u e^{i \mu_1 J_u Z} w e^{i \nu_1 J_w Z})^t]({\rm tr}[(u e^{i \mu_1 J_u Z} w e^{i \nu_1 J_w Z})^t])^*] =&\mathbb{E}[|{\rm tr}[(u w)^t]|^2] = \mathbb{E}[|{\rm tr}[u^t]|^2]  = {\rm min}(t,2),\label{eq:SFFSU2}\\
\notag\\
\mathbb{E}[{\rm tr}[(u e^{i \mu_1 J_u Z} w e^{i \nu_1 J_w Z})^t]({\rm tr}[(u e^{i \mu_1 J_u Z} w e^{-i \nu_1 J_w Z})^t])^*]=&\mathbb{E}[{\rm tr}[(u w)^t]({\rm tr}[(u w e^{-i 2 \nu_1 J_w Z})^t])^*] \notag\\
= & \mathbb{E}[{\rm tr}[u^t]({\rm tr}[(u e^{-i 2 \nu_1 J_w Z})^t])^*]\notag\\
= & \mathbb{E}[{\rm tr}[u^t]({\rm tr}[(u e^{-i 2 J_w Z})^t])^*],\\
\notag\\
\mathbb{E}[{\rm tr}[(u e^{i \mu_1 J_u Z} w e^{i \nu_1 J_w Z})^t]({\rm tr}[(u e^{-i \mu_1 J_u Z} w e^{i \nu_1 J_w Z})^t])^*] =&\mathbb{E}[{\rm tr}[(u w)^t]({\rm tr}[(u e^{-2 i \mu_1 J_u Z} w)^t])^*] \notag\\
= &\mathbb{E}[{\rm tr}[u^t]({\rm tr}[(u e^{-2 i \mu_1 J_u Z})^t])^*]\notag\\
= & \mathbb{E}[{\rm tr}[u^t]({\rm tr}[(u  e^{-i 2 J_u Z})^t])^*],\\
\notag\\
\mathbb{E}[{\rm tr}[(u e^{i \mu_1 J_u Z} w e^{i \nu_1 J_w Z})^t]({\rm tr}[(u e^{-i \mu_1 J_u Z} w e^{-i \nu_1 J_w Z})^t])^*] =&\mathbb{E}[{\rm tr}[u^t]({\rm tr}[(u w^\dag e^{-2 i \mu_1 J_u Z} w e^{-2 i \nu_1 J_w Z})^t])^*]\notag\\
=&\mathbb{E}[{\rm tr}[u^t]({\rm tr}[(u w^\dag e^{-2 i J_u Z} w e^{-2 i J_w Z})^t])^*]\notag\\
= & \mathbb{E}[{\rm tr}[u^t]({\rm tr}[(u e^{- i \alpha_w Z})^t])^*],
\end{align}
Here we used the well-known expression for the spectral form factor of the circular unitary ensemble~\cite{mehta2014random}, we exploited the invariance of the measure under 
\be
u \mapsto X u X
\ee
to remove $\mu_1$ and $\nu_1$, and we denoted by $e^{\pm i \alpha_w}$ the eigenvalues of $w^\dag e^{-2 i J_u Z} w e^{-2 i J_w Z}$. In particular, parametrising $w$ in terms of Euler angles ({cf}.~\eqref{eq:Eulervw}) we have 
\be
\alpha_w(b) \!= \!{\rm atan}\sqrt{\frac{1}{(\cos(J_u)\cos(J_w)- \cos(b) \sin(J_u)\sin(J_w))^2}-1}\in[|J_u-J_w|,J_u+J_w]\subset (0,\pi/2)\,.
\ee
The remaining averages are evaluated by expressing explicitly the Haar measure for ${\rm SU}(2)$ as follows 
\be
\mathbb{E}[f(u)]= \frac{1}{16\pi^2}  \int_{0}^{2\pi}  \!\!\!\!{\rm d}a \int_0^{\pi}   \!\!\!\!\!{\rm d}b\int_0^{4\pi} \!\!\!\!\!{\rm d}c\, \sin(b)  f(u(a,b,c))\,,
\ee
where $u(a,b,c)$ is the Euler angle parameterisation \eqref{eq:Eulervw}. This leads to 
\begin{align}
\mathbb{E}[{\rm tr}[(u e^{i \mu_1 J_u Z} w e^{i \nu_1 J_w Z})^t]({\rm tr}[(u e^{i \mu_1 J_u Z} w e^{-i \nu_1 J_w Z})^t])^*] =& f(t,2J_w),\label{eq:res2}\\
\mathbb{E}[{\rm tr}[(u e^{i \mu_1 J_u Z} w e^{i \nu_1 J_w Z})^t]({\rm tr}[(u e^{-i \mu_1 J_u Z} w e^{i \nu_1 J_w Z})^t])^*] =& f(t,2J_u),\label{eq:res3}\\
\mathbb{E}[{\rm tr}[(u e^{i \mu_1 J_u Z} w e^{i \nu_1 J_w Z})^t]({\rm tr}[(u e^{-i \mu_1 J_u Z} w e^{-i \nu_1 J_w Z})^t])^*] =&  g(t,J_u,J_w) ,
\label{eq:res4}
\end{align}
where we introduced the functions
\begin{align}
&f(x,a) \!=\! f_-(x,a)+f_+(x,a),\qquad  g(x,A,B)= \frac{1}{\sin(A)\sin(B)}\int_{|A-B|}^{A+B}   \!\!\!\!\!{\rm d}a\, \sin(a) f(x,a),\notag\\
&f_\pm(x,a) \!=\!  \frac{4}{\pi}  \int_{0}^{1} \!\!\!\!{\rm d}y\!\! \int_0^{\pi}  \!\!\!\!{\rm d}\varphi\, y \cos\!\!\left[x \!\left(\!\arctan\sqrt{\frac{1}{y^2\cos^2(\varphi)}-1} \pm \arctan\sqrt{\frac{1}{y^2\cos^2(\varphi - a)}-1}\right)\right] {\rm sign}\!\!\left[\frac{\cos(\varphi-a)^x}{\cos(\varphi)^x}\right]\!,
\end{align}
with first variable $x$ being an integer.
Plugging (\ref{eq:SFFSU2}, \ref{eq:res2}, \ref{eq:res3}, \ref{eq:res4}) back into \eqref{eq:Tabcd} we find 
\be
\!\!\mathbb B_1 \!=\!\! 
\left(\begin{array}{cccc}
 {\rm min}(t,2) +f(t,2J_u)+f(t,2J_w)+g(t,J_u,J_w) & 0 & 0 &  {\rm min}(t,2) +f(t,2J_u) - f(t,2J_w)-g(t,J_u,J_w) \\
0 & 0 &0 & 0  \\
0 & 0 &0 & 0 \\
{\rm min}(t,2) - f(t,2J_u)+f(t,2J_w)-g(t,J_u,J_w) & 0 & 0 &  {\rm min}(t,2) -f(t,2J_u)-f(t,2J_w)+g(t,J_u,J_w) 
\end{array}\right)\!\!.
\ee
This matrix has two non-zero eigenvalues reading as 
\begin{align}
&\!\!\!\bar \lambda_1(t) = {\rm min}(t,2)+g(t,J_u,J_w)+\sqrt{{\rm min}(t,2)^2+4 f(t,2J_u) f(t,2J_w)-2 g(t,J_u,J_w){\rm min}(t,2)+ g(t,J_u,J_w)^2})\,,\\
 &\!\!\!\lambda_1(t) = {\rm min}(t,2)+g(t,J_u,J_w)-\sqrt{{\rm min}(t,2)^2+4 f(t,2J_u) f(t,2J_w)-2 g(t,J_u,J_w){\rm min}(t,2)+ g(t,J_u,J_w)^2})\,.
\label{eq:lambda}
\end{align}
The asymptotic expansion of these eigenvalues for large times can be determined expanding the functions $f_\pm(x,a)$, and $g(x,J,J')$. This is done in Sec~\ref{sec:asy} and leads to   
\begin{align}
f(x,a) &\simeq 2 \cot(a) \frac{\sin(x a)}{x} + O\left(\frac{1}{x^{3/2}}\right)\!,\\
g(x,A,B)&\simeq  -\frac{2 \cos(a) \cos(xa)]}{x^2 \pi \sin(A)\sin(B)} \Bigr |_{|A-B|}^{A+B} + O\left(\frac{1}{x^{5/2}}\right)\!.
\end{align}
Note that in the limit $a\to0$ we have 
\be
\lim_{a\to0} f(x,a) = 2\,,
\ee
which recovers the large-time limit of \eqref{eq:SFFSU2} providing a useful consistency check. 

Plugging the asymptotic expansions into \eqref{eq:lambda} we find the expressions for $\lambda_\pm(t)$ and $K_1(t,L)$ given in the main text [Eq. (\ref{eq:exactK1haar})].

\subsection{Asymptotic expansion of $f_\pm(x,a)$ and $g(x,A,B)$}
\label{sec:asy}

In this appendix we determine the leading term in the asymptotic expansion of $f_\pm(x,a)$ for $a\in (0,\pi/2)$ and $g(x,A,B)$ with $A,B\in[0,\pi/4]$.  

Let us begin considering $f_\pm(x,a)$. To ease the manipulations we rewrite it in the following compact form 
\be
f_\pm(x,a)=  {\rm Re}\left[\iint\limits_{\mathcal D}{\rm d}y\,{\rm d}\varphi\, \chi(x,y,\varphi) e^{ i x\,  \Phi_\pm(y,\varphi) }\right]
\label{eq:integral}
\ee
where we introduced the domain $\mathcal D \equiv [0,1] \times [0,\pi]$ and the functions   
\be
\chi(x,y,\varphi)  = \frac{4}{\pi}y\, {\rm sign}\left[\frac{\cos(\varphi-a)^x}{\cos(\varphi)^x}\right],
\ee
and 
\be
\Phi_\pm(y,\varphi) := \arctan\sqrt{\frac{1}{y^2 \cos^2(\varphi)}-1} \pm \arctan\sqrt{\frac{1}{y^2 \cos^2(\varphi-a)}-1}.
\ee
Note that we included dependence on integer $x$ in $\chi(x,y,\varphi)$ to treat three different domains ($[0,1] \times [0,\pi/2]$, $[0,1] \times [\pi/2,\pi/2+a]$, and $[0,1] \times [\pi/2+a,\pi]$) at the same time.

Following the standard procedure~\cite{wong2001asymptotic} we compute the gradient of $\Phi_\pm$ to determine the critical points. A simple calculation yields 
\begin{align}
\partial_y \Phi_\pm(y,\varphi) &= - \frac{1}{\sqrt{\sec^2(\varphi)-y^2}}\mp\frac{1}{ \sqrt{{\sec^2(\varphi-a )}-y^2}}\,, \label{eq:grad1}\\
\partial_\varphi \Phi_\pm(y,\varphi) &=  \frac{y\tan(\varphi)}{\sqrt{\sec^2(\varphi)-y^2}}\pm\frac{y\tan(\varphi-a )}{ \sqrt{{\sec^2(\varphi-a )}-y^2}}\,. \label{eq:grad2}
\end{align}
From these expressions we see that the gradient of $\Phi_\pm$ does not vanish. To see it assume that \eqref{eq:grad1} vanishes for some $y=y_0\in[0,1]$ and $\varphi=\varphi_0\in[0,\pi]$. Then, plugging into \eqref{eq:grad2} we have  
\be
\partial_\varphi \Phi_\pm(y_0,\varphi_0) =  \pm\frac{y \cos(\varphi_0)}{\sqrt{1-y_0^2\cos^2(\varphi_0)}} (\tan (\varphi_0)-\tan (\varphi_0-a)), 
\ee
which never vanishes for $(y_0,\varphi_0)\in\mathcal D$ and $a\in (0,\pi/2)$. 

Since $\nabla \Phi_\pm\neq 0$ we can integrate ``by parts'' using the divergence theorem~\cite{wong2001asymptotic}. Namely, we use
\be
 \chi(x,y,\varphi) e^{i x \Phi_\pm(y,\varphi)} = \frac{1}{ix}\nabla\left(\frac{\nabla \Phi_\pm}{\|\nabla \Phi_\pm\|^2} \chi(x,y,\varphi) e^{i x \Phi_\pm(y,\varphi)}\right) - \frac{1}{ix}\nabla\left(\frac{\nabla \Phi_\pm}{\|\nabla \Phi_\pm\|^2} \chi(x,y,\varphi) \right) e^{i x \Phi_\pm(y,\varphi)}\,.
\ee  
Integrating over $\mathcal D$ we see that the first term becomes a boundary integral while the the second produces an integral of the same form as \eqref{eq:integral} but is suppressed by $x^{-1}$. Therefore we have 
\be
I_\pm = \iint\limits_{\mathcal D}{\rm d}y\,{\rm d}\varphi\, \chi(x,y,\varphi) e^{ i x\,  \Phi_\pm(y,\varphi) } = \frac{1}{ix}\int\limits_{\partial \mathcal D}{\rm d}s \, \frac{\boldsymbol n \nabla \Phi_\pm}{\|\nabla \Phi_\pm\|^2} \chi(x,y,\varphi) e^{i x \Phi_\pm(y,\varphi)} + O(I_\pm/x)
\ee
where $\partial\mathcal D$ is the boundary of $\mathcal D$ and $\boldsymbol n$ is its outward normal unit vector. Writing explicitly the boundary integral we have 
\begin{align}
\int\limits_{\partial \mathcal D}{\rm d}s \, \frac{\boldsymbol n \nabla \Phi_\pm}{\|\nabla \Phi_\pm\|^2} \chi(x,y,\varphi) e^{i x \Phi_\pm(y,\varphi)}  =&  \int_0^{\pi} {\rm d}\varphi \, \frac{\partial_y \Phi_\pm(1,\varphi)}{\|\nabla \Phi_\pm(1,\varphi)\|^2} \chi(x,1,\varphi) e^{i x \Phi_\pm(1,\varphi)}\notag\\
& + \int_0^1 {\rm d}y \, \frac{\partial_\varphi \Phi_\pm(y, \pi)}{\|\nabla \Phi_\pm(y, \pi)\|^2} \chi(x,y, \pi) e^{i x \Phi_\pm(y,\pi)}\notag\\
&- \int_0^1 {\rm d}y \, \frac{\partial_\varphi \Phi_\pm(y,0)}{\|\nabla \Phi_\pm(y,0)\|^2} \chi(x,y,0) e^{i x \Phi_\pm(y,0)} \notag\\
=& -\frac{4}{\pi}  \int_0^{\pi/2} {{\rm d} \varphi} \, \frac{(\cot(\varphi)\pm|\cot(\varphi-a)|) e^{ i x (\varphi\pm|\varphi-a|) }}{(\cot(\varphi)\pm|\cot(\varphi-a)|)^2+(1\pm {\rm sgn}(\varphi-a))^2} \notag\\
& +\frac{4}{\pi}  \int_{\pi/2}^{\pi/2+a} {{\rm d} \varphi} \, \frac{(\cot(\varphi)\mp \cot(\varphi-a) ) e^{ i x (-\varphi\pm (\varphi-a)) }}{(\cot(\varphi)\mp \cot(\varphi-a))^2+(1\mp 1)^2} \notag\\
& +\frac{4}{\pi}  \int_{\pi/2+a}^{\pi} {{\rm d} \varphi} \, \frac{(\cot(\varphi)\pm \cot(\varphi-a)) e^{-i x (\varphi\pm(\varphi-a)) }}{(\cot(\varphi)\pm \cot(\varphi-a))^2+(1\pm 1)^2} + O\left(\frac{1}{\sqrt x}\right)\!.
\end{align}
Here in the second step we used that $\Phi_\pm(y,0)$ and $\Phi_\pm(y,\pi)$ are non-trivial functions of $y$. Therefore, the integrals on the second and third line decay at least as $x^{-1/2}$.

Evaluating the elementary integrals and putting all together we finally obtain  
\begin{align}
f_-(x,a) &\simeq 2 \cot(a) (\pi - 2 a) \frac{\sin(x a)}{\pi x} + O\left(\frac{1}{x^{3/2}}\right), \\
f_+(x,a)&\simeq 4 a \cot(a)\frac{\sin(x a)}{\pi x} +O\left(\frac{1}{x^{3/2}}\right)\,.
\end{align}

Let us now briefly consider $g(x,A,B)$. Using the short-hand notation introduced above we have 
\be
g(x,A,B)= \frac{1}{\sin(A)\sin(B)} \sum_{\mu=\pm} {\rm Re}\left[\iiint\limits_{\mathcal F}{\rm d}a\,{\rm d}y\,{\rm d}\varphi\, \sin(a)\chi(x,y,\varphi) e^{ i x\,  \Phi_\mu(y,\varphi,a) }\right]
\ee
where we introduced the explicit dependence on $a$ and denoted by $\mathcal F$ the integration domain $[|A-B|,A+B] \times \mathcal D$. 

Since $\nabla \Phi_\mu(y,\varphi,a)\neq 0$ in $\mathcal F$ we can again integrate by parts and consider only the boundary integrals. This leads to 
\begin{align}
g(x,A,B)\simeq \frac{1}{x \sin(A)\sin(B)}  \sum_{\mu=\pm} {\rm Im}&\left[\iint\limits_{\mathcal D}{\rm d}y{\rm d}\varphi \, \chi(x,y,\varphi) \frac{ \sin(a) \partial_a \Phi_\mu}{\|\nabla \Phi_\mu\|^2}  e^{i x \Phi_\mu(y,\varphi,a)}\Big |_{a=A_-}^{A_+} \right.  \notag\\
&+ \iint\limits_{\mathcal D_1}{\rm d}a{\rm d}y \, \chi(x,y,\varphi) \frac{ \sin(a) \partial_\varphi \Phi_\mu}{\|\nabla \Phi_\mu\|^2}  e^{i x \Phi_\mu(y,\varphi,a)}\Big |_{\varphi=0}^{\pi/2} \notag\\
&\left.+ \iint\limits_{\mathcal D_2}{\rm d}a{\rm d}\varphi \, \chi(x,1,\varphi) \frac{ \sin(a) \partial_y \Phi_\mu}{\|\nabla \Phi_\mu\|^2}  e^{i x \Phi_\mu(1,\varphi,a)}\right],
\end{align}
where we set $A_\pm=|A\pm B|$, $\mathcal D_1=[A_-,A_+] \times [0,1]$, $\mathcal D_2=[A_-,A_+] \times [0,\pi]$ and 
\be
\partial_a \Phi_\pm(y,\varphi,a) = \mp\frac{y \tan(\varphi-a )}{ \sqrt{{\sec^2(\varphi-a )}-y^2}}\,.
\ee
Integrating the first two terms again by parts we find the following leading contributions coming from first and last term
\begin{align}
g(x,A,B)\simeq & \frac{4\sin a \cos(ax)}{\pi x^2 \sin(A)\sin(B)} \int_{0}^{\pi}\!\!\!{\rm d}\varphi \frac{1}{(\cot(\varphi)-\cot(\varphi-a)}\Big |_{a=A_-}^{A_+} = -\frac{2\cos(ax)  \cos(a)}{x^2 \sin(A)\sin(B)} \Big |_{a=A_-}^{A_+}\,. 
\end{align}

\section{Infinite-time limit of $\mathbb B_n$}
\label{sec:barBn}
Plugging \eqref{eq:infinitetimeav} into \eqref{eq:Tabcd} we find 
\be
\lim_{t\rightarrow \infty} [\mathbb B_n]_{a,b} = \sum_{\bf x}  g_{\bf x} \prod_{j=1}^{2n}\!\left[\sum_{\mu,\nu=\pm} \mu^{[a]_j} \nu^{[b]_j} e^{-ix_{\mu\nu}}\!\right]\!\!, 
\label{eq:Bntinftyapp}
\ee
where 
\be
{\bf x} = (x_{++}, x_{+-}, x_{-+}, x_{--}),\qquad\qquad \sum_{\bf x}=\sum_{x_{++}}\sum_{x_{-+}}\sum_{x_{+-}}\sum_{x_{--}}
\ee
$x_{\mu\nu}=2\pi n_{\mu\nu}/(2n+1)$ with $n_{\mu\nu}=0,\ldots,2n$,
\be
g_{\bf x}=\prod_{\mu,\nu=\pm} g_{x_{\mu\nu}},\qquad\qquad
g_x= \frac{1}{(2n+1)}\sum_{k=0}^n\frac{e^{i 2 k x}}{4^{k}}\binom{2k}{k}.
\ee
Using the expression \eqref{eq:Bntinftyapp} we find 
\be
\lim_{t\rightarrow \infty} {\rm tr}[{\mathbb B}_n^k]= 2^{2 n k} \sum_{\bf x_1} \cdots  \sum_{\bf x_k} {\rm tr}[M({\bf x}_1)\cdots M({\bf x}_k)]^{2n} g_{{\bf x}_1}\cdots g_{{\bf x}_k}
\label{eq:simplification}
\ee
where we set 
\be
M({\bf x}) = \begin{bmatrix}
e^{-i x_{++}} & e^{-i x_{+-}} \\
e^{-i x_{-+}} & e^{-i x_{--}} \\ 
\end{bmatrix}\,. 
\ee
Eq.~\eqref{eq:simplification} implies that $\lim_{t\to\infty}{\mathbb B}_n$ has the same spectrum as 
\be
\bar{\mathbb B}_n =  2^{2 n} \sum_{\bf x} M({\bf x})^{\otimes 2n} g_{{\bf x}}
\ee
Defining 
\be
N_{\mu\nu}(a,b)= \sum_{j=1}^{2 n} \delta_{\mu,a_j}  \delta_{\nu,b_j},\qquad\qquad B(k)= \begin{cases}
\displaystyle \,\,\,\,\,\,\,0 & {\rm mod}(k,2) =1\\
&\\
\displaystyle \binom{k}{k/2} & {\rm mod}(k,2)=0 
\end{cases}, 
\ee
and using 
\be
\sum_{x} e^{- i k x }g_x = \frac{1}{2^{2k}} B(k)
\ee
we find that $\bar{\mathbb B}_n$ has the following matrix elements
\be
[\bar{\mathbb B}_n]_{a,b}= B(N_{++}(a,b))B(N_{-+}(a,b))B(N_{+-}(a,b))B(N_{--}(a,b))\,,
\ee
which are those given in Eq.~\eqref{eq:Bntinfty} of the main text. 

\section{Explicit form of $r_n(\xi)$}
\label{sec:rn}

Computing the moments of the spectral form factor for the time evolution operator \eqref{eq:Upatch} ($\xi$ is assumed to divide $2L$) we find 
\be
K_{n,\xi}(L) = \mathbb{E}\left[ |{\rm tr}[\mathbb U_{\xi,L}^t]|^{2n} \right]= r_n({\xi})^{2L},
\ee
where $\mathbb{E}\left[\cdot\right]$ is the average over flat distribution of eigenphases of $u^{(\xi)}$, and is hence clearly independent of $t\neq 0$, and we defined 
\be
r_n({\xi}) : = \left(\mathbb{E}\left[ |{\rm tr}[u^{(\xi)}]|^{2n} \right]\right)^{1/\xi}\,.
\ee
Therefore, $r_n({\xi})$ can be expressed in terms of the $n$-th moment of the SFF of a $2^\xi$ dimensional unitary Poisson random matrix. The latter can be expressed in terms of the generating function~\cite{haake2018}
\be
G(x) = J_0(x)^{2^\xi},
\ee
where $J_n(x)$ is the Bessel function of the first kind, as follows
\be
\mathbb{E}\left[ |{\rm tr}[u^{(\xi)}]|^{2n} \right] = \left(- \frac{1}{x}\partial_x x \partial_x\right)^n G(x)\big|_{x=0}\,.
\ee 
This gives
\begin{align}
r_1(\xi) & = 2,\\
r_2(\xi) & = ((2!)\cdot 2^{2\xi}-2^{\xi})^{1/\xi},\\
r_3(\xi) & = ((3!)\cdot 2^{3\xi}-9\cdot 2^{2\xi}+5\cdot 2^{\xi})^{1/\xi},\\
r_4(\xi) & = ((4!)\cdot 2^{4\xi} - 72\cdot 2^{3\xi} + 82\cdot 2^{2\xi} - 33\cdot 2^{\xi})^{1/\xi},\\
r_5(\xi) & = ((5!)\cdot 2^{5\xi}- 600\cdot 2^{4\xi}+ 1250\cdot 2^{3\xi} - 1225 \cdot 2^{2\xi}+456 \cdot 2^{\xi})^{1/\xi},\\
r_6(\xi) & = ((6!)\cdot 2^{6\xi}- 5400\cdot 2^{5\xi}+ 17700\cdot 2^{4\xi}- 30600\cdot 2^{3\xi}+ 27041\cdot 2^{2\xi}-9460\cdot 2^{\xi})^{1/\xi}.
\end{align}
In the particular case $\xi=1$ the generic moment is expressed as 
\be
r_n(1) = B(2n) = \binom{2n}{n}\,.
\ee

\section{Further numerical data}
\label{sec:numerics}

\begin{figure}[t!]
    \centering
   \includegraphics[width=0.5\textwidth]{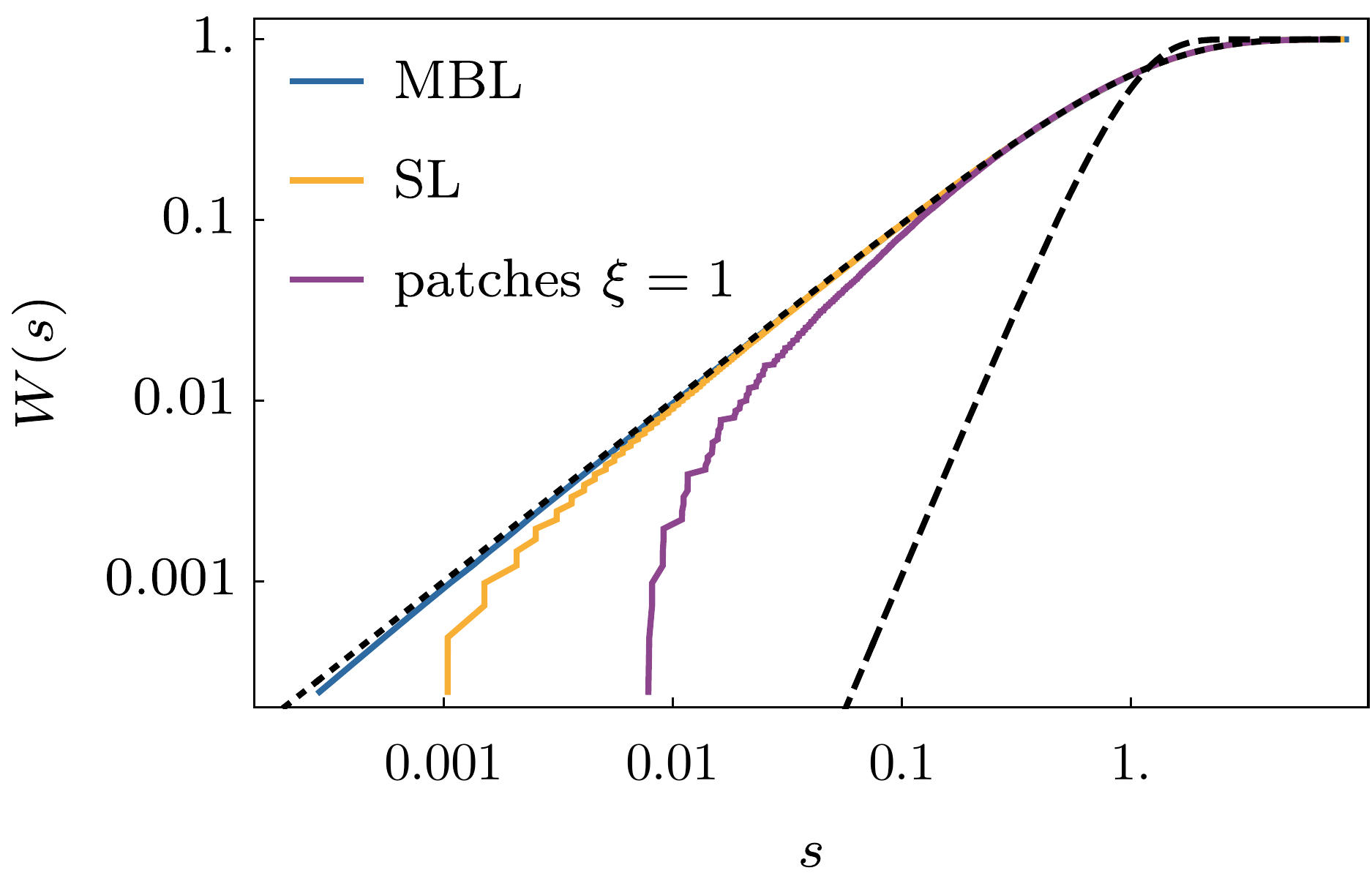}
    \caption{
	Cumulative nearest-neighbour level spacing distributions, $W(s)$ is the probability of a random spacing being smaller than $s$, for the MBL regime at $J=0.01$ (blue), strongly localisated at $J=0.5$ (yellow), and independent random spins (magenta). We take $2L=12$ and average over 1000 disorder realizations. Black dotted and dashed lines show predictions for Poisson and CUE ensembles, respectively.
	}
    \label{fig:levelspacings}
\end{figure}

Here we provide some additional numerical data. Firstly, in Fig.~\ref{fig:levelspacings} we show cumulative level spacing distributions of generic local Floquet circuits in MBL regime~\eqref{eq:MBLcircuit}, strongly localised systems~\eqref{eq:Floquetgates}, and systems formed from disconnected patches~\eqref{eq:Upatch}. As we have seen from the moments of SFF in Eqs.~\eqref{eq:momentstinfty} and \eqref{eq:independentspins}, the spectral statistics for the latter two models are not Poissonian, which influences the level spacing distributions only  at extremely short distances (of a tiny fraction of mean level spacing). It is important to note that the discrepancies are much more apparent when looking at the higher moments of SFF.

\begin{figure}
    \centering
   \includegraphics[width=0.49\textwidth]{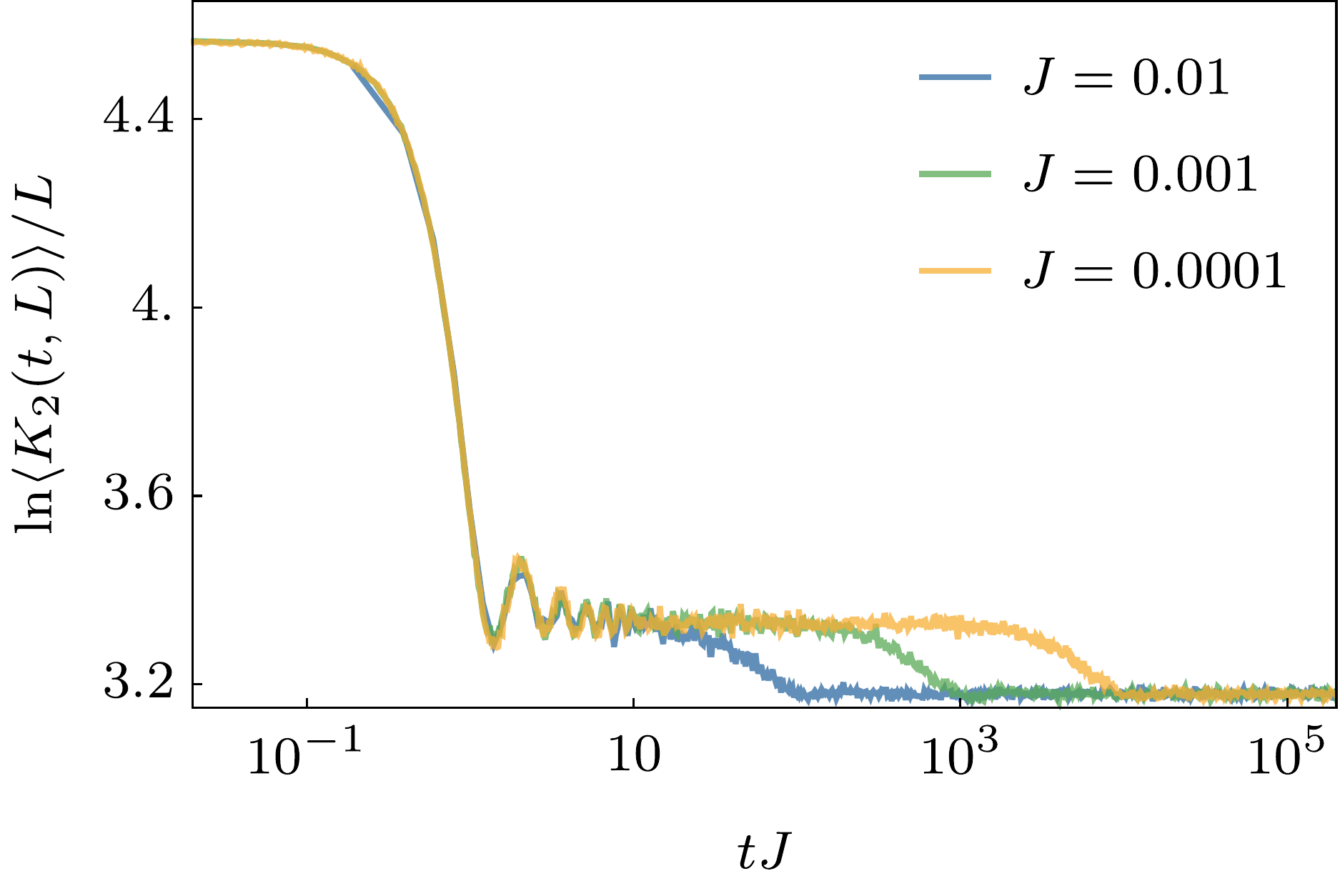}
\includegraphics[width=0.49\textwidth]{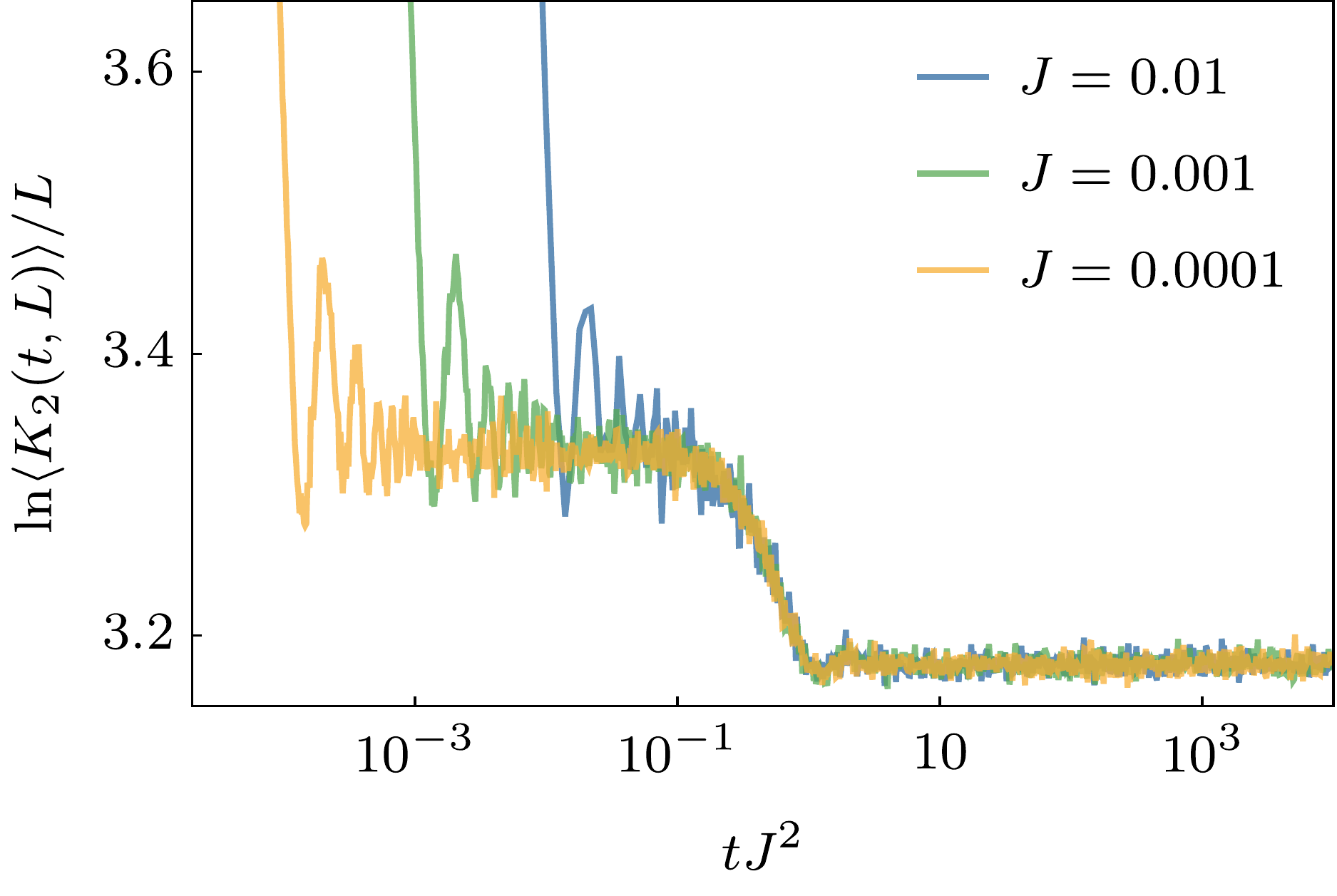}
    \caption{
	 Second moment of the SFF for the strongly localised circuits \eqref{eq:Floquetgates}, versus $t J$ and $t J^2$, respectively. We see a nice collapse of the data indicating scalings of crossover times $\tau_1\propto J^{-1}$ and $\tau_2\propto J^{-2}$.
Here we took $2L=10$, different coloured lines correspond to the three values of $J$, and we averaged over $200000$ realisations ($10000$  independently drawn disorder samples and windows of $20$ nearby time steps). }
    \label{fig:compareJsSL}
\end{figure}

\begin{figure}
    \centering
   \includegraphics[width=0.49\textwidth]{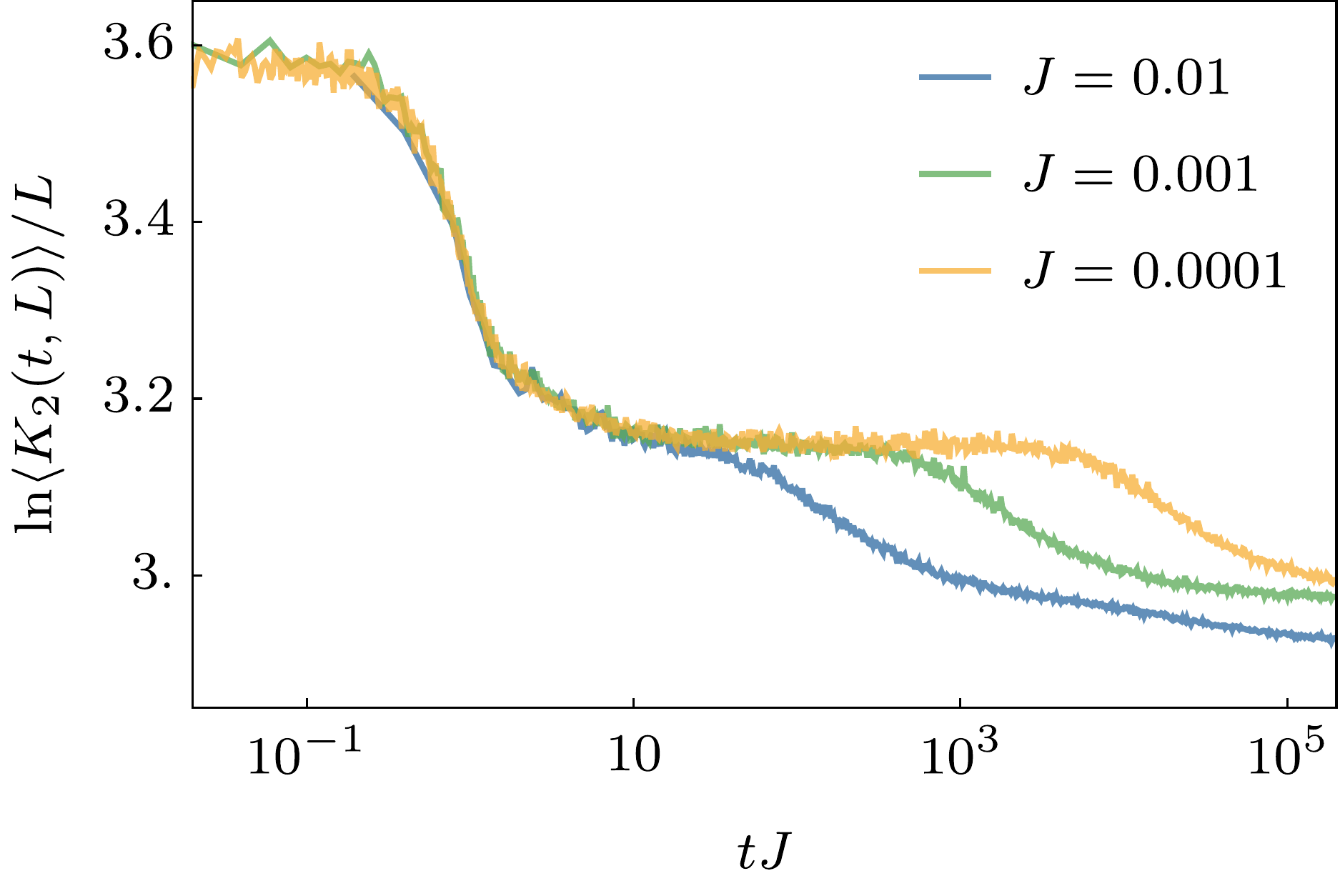}
\includegraphics[width=0.49\textwidth]{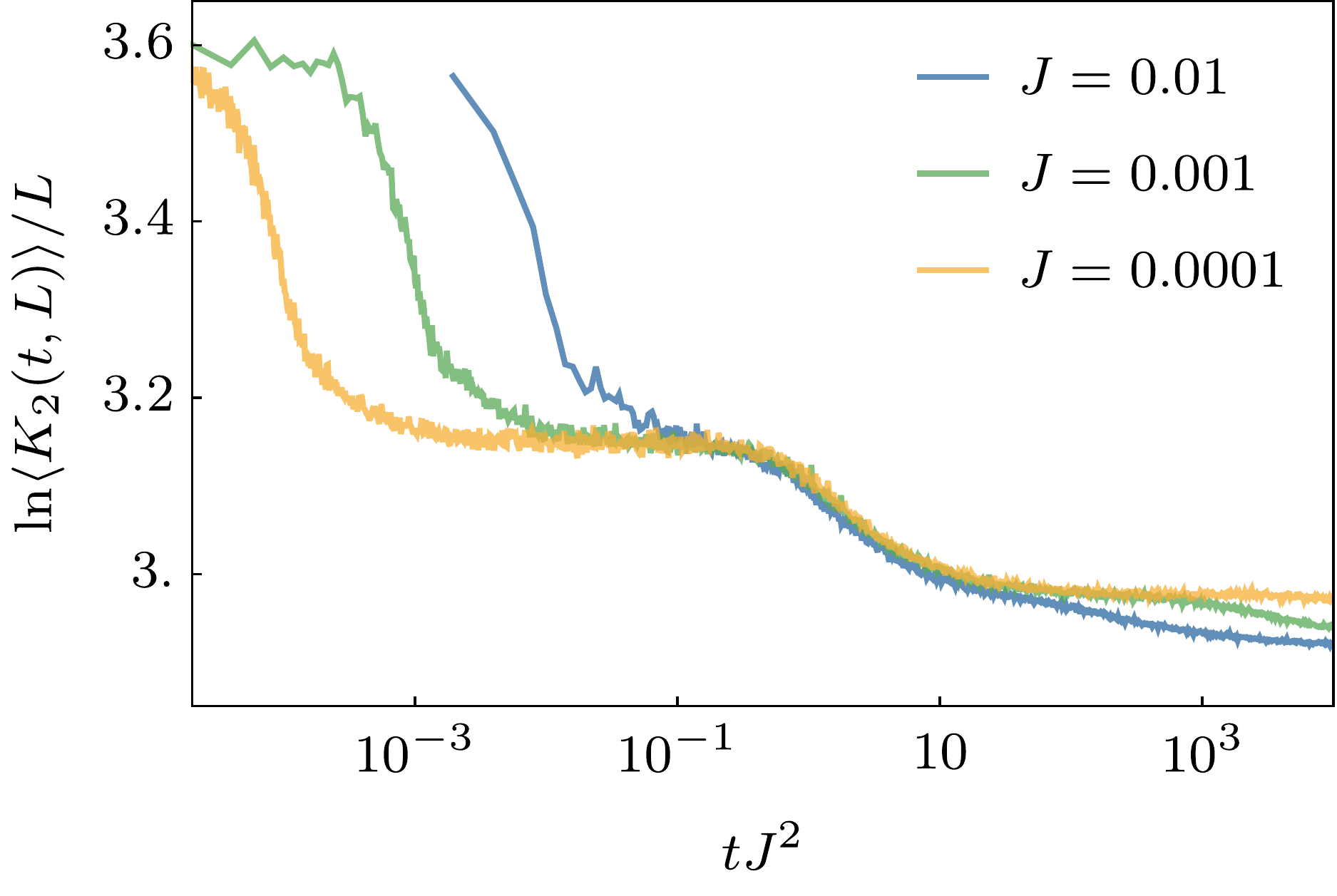}
    \caption{
	 Second moment of the SFF for the circuits \eqref{eq:MBLcircuit}, versus $t J$ and $t J^2$, respectively. We see a nice collapse of the data indicating scalings of crossover times $\tau_1\propto J^{-1}$ and $\tau_2\propto J^{-2}$.
Here we took $2L=10$, different coloured lines correspond to the three values of $J$, and we averaged over $200000$ realisations ($10000$  independently drawn disorder samples and windows of $20$ nearby time steps).	}
    \label{fig:compareJsMBL}
\end{figure}

Secondly, we demonstrate the scaling of the crossover times claimed in the main text, i.e. 
\be
\tau_1\propto J^{-1},\qquad \tau_2\propto J^{-2},
\ee
for both circuits, Eq.~\eqref{eq:Floquetgates} with $J_u=J_w=J$, and more generic ones, Eq.~\eqref{eq:MBLcircuit}. This is evident from the data collapses shown in Figs.~\ref{fig:compareJsSL} and \ref{fig:compareJsMBL}.

Thirdly, we compare data for four different generic MBL models. The first model is given by Eq.~\eqref{eq:MBLcircuit}. The second is the disordered kicked Ising (KI) model given in terms of the Floquet operator:
\be
\mathbb U = e^{i \sum_{j \in \frac{1}{2}\mathbb Z_{2L}}   g X_j} e^{i \sum_{j \in \frac{1}{2}\mathbb Z_{2L}} J Z_j Z_{j+1/2}+ h_j Z_j} \, ,
\label{eq:KIcircuit}
\ee
where $h_j$ are random variables and we used the short hand notation $O_j = \eta_{j,L}(O)$.
The third model couples also the next-nearest neighbours (NNN):
\begin{align}
\displaystyle \mathbb U &= \mathbb U_{d1} e^{i J \sum_{j =0,3/2,6/2,...}Q_j } 
\mathbb U_{d2} e^{i J \sum_{j=1/2,4/2,7/2,...}Q_j} 
\mathbb U_{d3} e^{i J \sum_{j =2/2,5/2,8/2,...}Q_j }\,, \\
Q_j&=Z_j Z_{j+1}+ (Z_j Z_{j+1/2}+Z_{j+1/2} Z_{j+1})/2 \,, 
\label{eq:NNNcircuit}
\end{align}
where
\begin{equation}
\mathbb U_{dk}= \bigotimes_{x\in\frac{1}{2}\mathbb Z_{2L}}u_{x,k}
\end{equation} 
are disorder layers made out of tensor products of random independent SU(2) matrices $u_{x,k}$. 
The fourth, Floquet $XYZ$ model is given by:
\begin{align}
\displaystyle \mathbb U &= \mathbb U_{d1} e^{i  \sum_{j =0,1,...}Q'_j } 
\mathbb U_{d2} e^{i  \sum_{j=1/2,3/2,...}Q'_j} , \\
Q'_j&=J_x X_j X_{j+1/2}+ J_y Y_j Y_{j+1/2}+ J_z Z_j Z_{j+1/2} \, .
\label{eq:XYZ}
\end{align}
Notice that it breaks $\mathbb{Z}_2$ symmetry.

\begin{figure}[t!]
    \centering
   \includegraphics[width=0.55\textwidth]{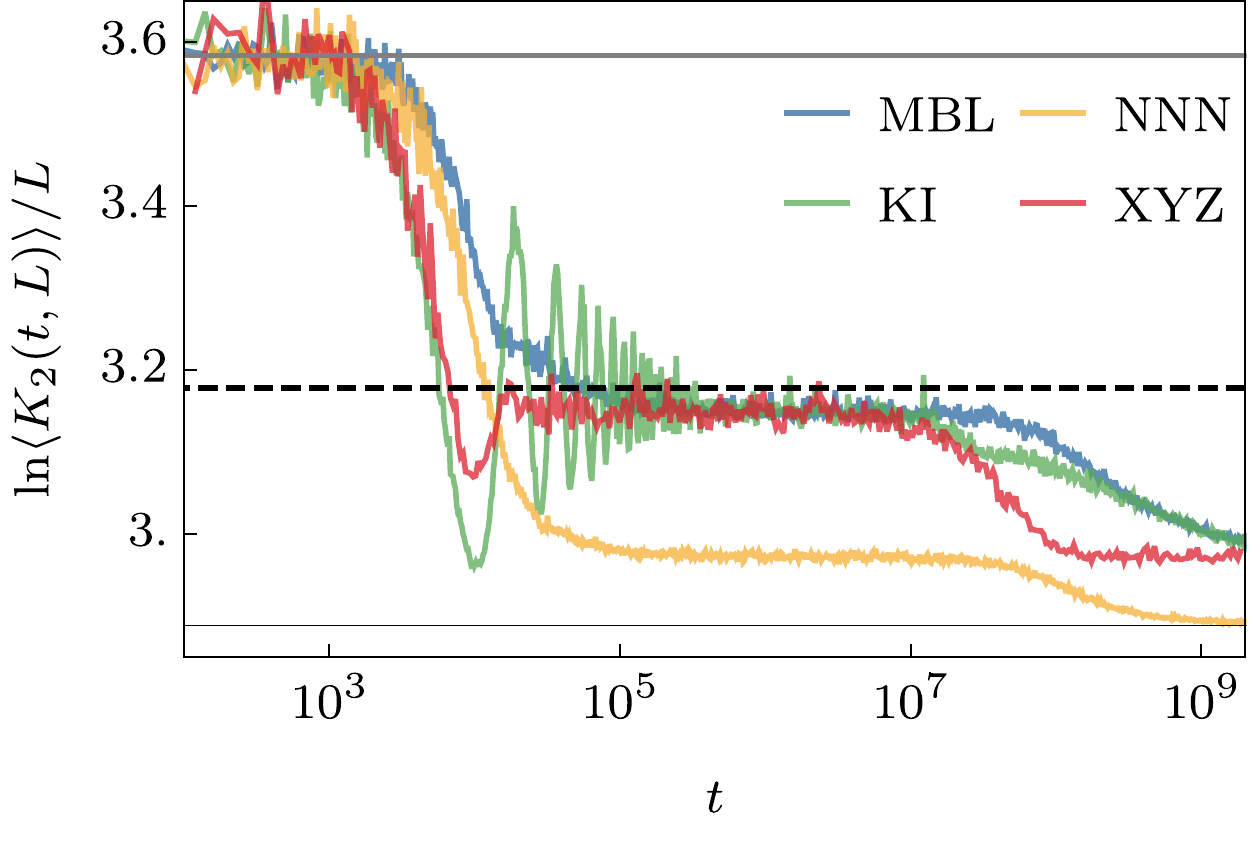}
    \caption{
	 Second moment of the SFF for the circuits \eqref{eq:MBLcircuit} (MBL), \eqref{eq:KIcircuit}(KI), \eqref{eq:NNNcircuit}(NNN) and \eqref{eq:XYZ}(XYZ) versus time. Here we took $J=10^{-4}$, different coloured lines correspond to the four different models, and we averaged over $140000$ realisations ($7000$ independently drawn samples and $20$ nearby time steps). Dashed black and solid grey lines indicate respectively the result \eqref{eq:momentstinfty} and  \eqref{eq:independentspins} for $\xi=1$. Thin black line indicates the Poisson result for $2L=12$, which is approched by all models at very long times (not shown). Data is for $2L=12$, and for the second model shown in green we took $g=0.5$.
	 For the fourth model (XYZ) we took $J_x=J, J_Y=0.5 J, J_Z= \pm 1.5J$, with + in the first half layer and - in the second, and averaged over $1700$ independently drawn disorder samples and windows of $40$ nearby time steps.
	 }
    \label{fig:compareMBLs}
\end{figure}

We compare these four examples in Fig. \ref{fig:compareMBLs}, where we see similar qualitative behaviour of the fluctuations of SFF. In particular, we clearly see the cascade of the three (four in the case of $XYZ$ model) regimes discussed in the main text. The value of the middle plateau is different for the NNN model, which illustrates that the value is model-dependent.

\twocolumngrid

\bibliography{bibliography}

\end{document}